\documentclass[onecolumn,preprintnumbers,amsmath,amssymb,aps,prd]{revtex4-1}
\usepackage{bm,tikz}
\usepackage{color}

\usepackage{graphicx,epsfig,epsf,color}
\usepackage{multirow}
\usepackage{float}
\usepackage[utf8x]{inputenc}
\usepackage{amssymb,amsmath,appendix}
\usepackage{slashed}
\usepackage{subfigure}
\usepackage{pict2e}
\usepackage{array}
\usepackage[colorlinks=true,citecolor=blue,urlcolor=blue,linkcolor=blue]{hyperref}
\usepackage{ulem}

\begin{document}
\title{Role of density profile of sub-GeV dark matter in the properties of dark matter admixed quark stars with Bayesian analysis of dark-NJL model}

\author{Debashree Sen$^1$}
\email{debashreesen88@gmail.com}

\author{Atanu Guha$^2$}
\email{atanu@cnu.ac.kr}

\author{Jong-Chul Park$^{2}$}
\email{jcpark@cnu.ac.kr}

\affiliation{$^1$Department of Physics Education, Daegu University, Gyeongsan 38453, South Korea}
\affiliation{$^2$Department of Physics, Chungnam National University,\\ 99, Daehak-ro, Yuseong-gu, Daejeon-34134, South Korea}

\date{\today}




\begin{abstract}

We investigate the structural and oscillation properties of dark matter (DM) admixed strange quark stars (DMSQSs). The strange quark matter (SQM) is described with the well-known Nambu-Jona-Lasino (NJL) model and the self-interacting fermionic DM is included in a systematic manner. The self-interaction of DM is of four-Fermi type and the overall DM density is considered as a function of the baryon density of SQM with two free parameters ($\alpha$, $\rho_{sc}$). This work is the first to consider four-Fermi interactions between fermionic DM and SQM in DMSQSs. Certain experiments like LZ, XENON, DarkSide, CRESST, and LHC have almost ruled out the possibility of contact interaction between SQM and massive DM (in GeV order). Recently, the quest for sub-GeV DM has garnered significant attention. We show that recent astrophysical constraints on the structural properties of compact stars also do not support the presence of massive DM in DMSQSs. On the other hand, we find sub-GeV DM to successfully concur with such observational constraints. We also calculate the fundamental $f$-mode frequency ($f_f$) of the DMSQSs, which shows universality with compactness, mean density, and tidal deformability. Further, we investigate the prospect of detection of $f_f$ with respect to the projected sensitivity of upcoming gravitational wave detectors like aLIGO, A+, Cosmic Explorer, and Einstein Telescope. In our DMSQS model, the three free parameters are $\alpha$, $\rho_{sc}$, and the ratio of repulsive to attractive interaction in SQM ($G_V/G_S$), which are optimized by Bayesian analysis in light of various recent astrophysical data.

\end{abstract}




\maketitle



\section{Introduction}
\label{sec:Introduction}

Compact star physics is one of the most inconclusive and at the same time very interesting topics of research. The extreme conditions of density (about $5-10$ times normal nuclear density), compactness, and gravity make it challenging to understand their composition and equation of state (EoS) from experimental perspectives. However, various theoretical modeling of compact star matter help us to understand the composition, interactions, and EoS of compact stars. Fortunately, the recent astrophysical observations of the structural properties like the mass ($M$), radius ($R$) and tidal deformability ($\Lambda$) of compact stars provide us with a scope to constrain the high density EoS of compact stars to some extent. Such constraints include the observational data obtained from the most massive pulsar PSR J0740+6620 \cite{Fonseca:2021wxt, Miller:2021qha, Riley:2021pdl}, GW170817 \cite{LIGOScientific:2018cki}, NICER data for PSR J0030+0451 \cite{Riley:2019yda, Miller:2019cac}, PSR J0437-4715 \cite{Choudhury:2024xbk}, and PSR J1231-1411 \cite{Salmi:2024bss}, and HESS J1731-347 \cite{Doroshenko:2022}. 

Several theoretical studies predicted the possible existence of strange quark stars (SQSs), composed of $u, d$ and $s$ quarks, collectively known as strange quark matter (SQM) \cite{Olinto:1986je}. Consequently, over many years, the theoretical modeling of SQM has been a topic of great interest to study the properties of SQSs. Such important models include the Nambu-Jona-Lasinio (NJL) model \cite{Nambu:1961tp,Nambu:1961fr}, MIT Bag model \cite{Chodos:1974je}, quark-mass density dependent model \cite{Lugones:2022upj}, the quasi-particle model \cite{Li:2019akk}, the
Dyson-Schwinger model \cite{Luo:2019dpm}, and the Polyakov Chiral SU(3) quark mean ﬁeld model \cite{Kumari:2021tik}. In the NJL model, the interactions between the quarks are described by contact-type four-Fermi operators. The model includes the scalar, pseudo-scalar, vector, and the t’Hooft six-fermion interaction \cite{Hatsuda:1987pc,Vogl:1991qt,Klevansky:1992qe,Hatsuda:1994pi,Kitazawa:2002jop,Lopes:2020rqn,Masuda:2012ed,Buballa:2003qv,Guha:2024gfe}. This model is popularly employed to study the properties of SQSs in \cite{Lopes:2020rqn,Masuda:2012ed,Guha:2024gfe}. In such works, the coupling related to the vector channel is denoted by $G_V$. It is seen that for a suitable choice of $G_V$, consistent with lattice QCD (LQCD) results, the constraints on the structural properties of compact stars are not satisfied. Therefore, in the present work we invoke the interaction between SQM and fermionic dark matter (DM), and study the properties of DM admixed SQSs (DMSQSs). We therefore name this model the dark-NJL (DNJL) model. Several works have been dedicated to studying the properties of DMSQSs with several models, considering SQM and DM to be both interacting and non-interacting \cite{Yang:2024sxi, Rezaei:2023iif, Panotopoulos:2017eig, Zheng:2016ygg, Mukhopadhyay:2015xhs, Jimenez:2021nmr, Yang:2021bpe, MafaTakisa:2020avv, Panotopoulos:2018ipq, Panotopoulos:2018joc, Cassing:2022tnn, Ferreira:2022fjo, Chanda:2025wgh, Panotopoulos:2025ixx, Liu:2025vwm, Jyothilakshmi:2024xtl, Zhen:2024xjc, Sen:2022pfr, Lopes:2023uxi, Lopes:2025jyz}. The present work is the first in literature to consider DM-SQM interaction via contact type four-Fermi operators in DMSQSs. For the SQM we employ the NJL model, which is well described in \cite{Lopes:2020rqn,Masuda:2012ed,Guha:2024gfe}.To describe the dark sector with the same footing as the SQM, we also consider the DM self-interaction via contact type four-Fermi operators. 

In our earlier works, we have considered a constant DM density in DM admixed compact stars \cite{Guha:2021njn,Sen:2021wev,Sen:2022pfr,Guha:2024pnn,Sen:2024yim,Jyothilakshmi:2024xtl} following \cite{Panotopoulos:2017idn} and many other works. However, such an assumption of the DM density profile may oversimplify the effects of DM in compact stars, and it is suggested that gravitational effects should lead to a
variable DM density profile in compact stars \cite{Kumar:2025ytm}. The intense gravitational field of compact stars may accrete DM from its surroundings and as the density of the core increases, the accreted DM tends to accumulate more towards the core. Therefore, the DM density is supposed to increase from the surface to the core. Studies, focusing on possible DM production in compact stars via dark decay of neutrons, also support this fact \cite{Baym:2018ljz,Husain:2022brl}. Moreover, the assumption of the constant DM density (DM Fermi momentum) being constant with the density (radius) of the star introduces thermodynamic inconsistency to the system \cite{Hajkarim_2025}. Therefore, another salient feature of the present work is that we consider the DM number density as a function of the baryon density of SQM in terms of two free parameters ($\alpha$, $\rho_{sc}$). This feature makes the DM density variable with the density of the star and helps us to overcome both the issues of oversimplification and thermodynamic inconsistency. Such a connection between the DM number density and the baryonic density of SQM has not been considered before in the literature. 

In spite of the strong indication regarding the existence of DM in our Universe, the interaction type and strength between DM and standard model (SM) particles are inconclusive till date. Over the last decade, various search avenues for DM (e.g., direct detection, indirect detection, and collider searches) excluded a vast region of the DM-SM interaction parameter space. Void detection results pushed us to expand the domain of DM searches. In particular, low-mass DM of sub-GeV order has been of growing interest in recent times~\cite{LZ:2025iaw, PandaX-II:2021kai, LUX:2018akb, Guha:2024mjr, Leane:2025efj, Dutta:2024kuj, Lee:2024wzd, Kim:2024ltz, Alhazmi:2025nvt, Mishra:2025juk}. Since we have considered contact-type of DM-DM and DM-SQM interaction in the present work, in the absence of any mediator, it is important to choose the DM mass ($m_{\chi}$) and the corresponding momentum cut-offs wisely, according to the results obtained from various experiments dedicated to the search for DM. In the context of the contact-type of interaction between DM and SQM, several direct detection experiments such as LZ \cite{LZ:2022lsv}, XENON \cite{XENON:2019rxp}, DarkSide \cite{DarkSide:2018bpj}, CRESST \cite{CRESST:2019jnq}, and the collider experiment LHC \cite{Roy:2024ear} have already ruled out most of the parameter space (effective field theory (EFT) expansion scale vs mass of DM) in the heavy DM mass ($m_{\chi} \geq 1~\rm{GeV}$) regime. The region allowed by these experiments for $m_{\chi} \geq 1~\rm{GeV}$ does not satisfy the relic bound. Therefore, in the present work, we consider very conservative benchmark values for $m_{\chi}$, compatible with these search experiment results and non-baryonic relic density constraint in the sub-GeV scale of $m_{\chi}$. The corresponding momentum cut-off for the DM self-interaction has been calculated using the existing DM self-interaction bound \cite{Tulin:2013teo, Tulin:2017ara} and the same for the DM-SQM interaction is obtained by satisfying the relic bound \cite{Belanger:2013oya}. Such parameterization of DM and its interaction with SQM is not performed before in literature, and therefore we name it SGP parameterization.

Deploying our DNJL model with the SGP parameterization, we obtain the DNJL EoS, and consequently calculate the structural properties of the DMSQSs. We next obtain the non-radial oscillation frequencies of the DMSQSs. Decades of observational and theoretical efforts indicate the strong possibility of the existence of several oscillation modes in compact stars. They are the fundamental ($f$), pressure ($p$), rotational ($r$), spacetime ($w$) and gravity ($g$) modes with corresponding frequencies \cite{Kokkotas:1999bd}. As per the present understanding, these oscillation modes can be initiated in various ways, e.g., abrupt variation of the star's internal structure like density discontinuity, rapid rotation, accretion of matter from a binary companion, interactions with external forces, nutation of the star, star-quakes etc. Several attempts are being made, and in the near future the detection of these oscillation modes is highly anticipated. The manifestation of oscillations is expected through the emission of different types of waves, such as gravitational waves (GWs) or electromagnetic radiation \cite{Belloni:2020cjs}. In the era of GW observation, the detection and study of such oscillation modes, at least the $f$-mode, is highly interesting and important. Several upcoming detectors like Advanced LIGO (aLIGO) \cite{KAGRA:2013rdx}, A+ \cite{A+}, Cosmic Explorer (CE1) \cite{CE1}, and Einstein Telescope (ET) \cite{Hild:2010id, Punturo:2010zz} are expected to detect the oscillation frequencies, which will open up a new way of enriching our knowledge regarding compact star interior. The oscillation of compact stars coupled with GW radiation is first studied by \cite{Thorne:1967apj, Thorne:1969rba}. In this work, we focus on the study of the non-radial oscillations of the DMSQS, which are the most general type of oscillations of stars. The stars deviate from their spherical shape in this scenario. The estimation of the oscillation frequency in a complete general relativistic (GR) approach is done by including both the perturbation of the matter inside the star and the perturbation of the spacetime metric due to the non-radial oscillation. The consolidated perturbation in general can be decomposed into spherical harmonics which contain both even and odd parity components. The odd parity perturbations are important only for rotating stars and manifested as $r$-mode of oscillations. For the non-rotating consideration, the odd parity perturbation gives trivial zero mode and only the even parity perturbations are relevant. In this work, we investigate only the even parity perturbation of the Regge-Wheeler metric \cite{Zhao:2022tcw} considering the non-rotating scenario. Later, \cite{Sotani:2010mx} simplified the derivation of the integrated solution with the help of Cowling approximations \cite{Cowling:1941nqk}, neglecting the metric perturbations. In the present work, we compute the $f$-mode oscillation frequency ($f_f$) of the DMSQSs with both the Cowling approximation and the full GR approach. The universality of $f_f$ is studied with respect to $M$, $R$, and $\Lambda$ and compared with other works. We also investigate the prospect of detection of such GW waves originating from the non-radial oscillations of the DMSQSs in terms of the amplitude of GW strain. We compare our results with the projected sensitivities of upcoming GW detectors such as aLIGO, A+, CE1, and ET.

To begin with, our consolidated DNJL model consists of three free parameters, viz., $G_V$, $\alpha$, $\rho_{sc}$. For several combinations of them, we obtain the structural properties of DMSQSs. We compare our results with the corresponding observational data in order to constrain the free parameters. From the initial rough estimations, we set the priors and then perform a proper Bayesian analysis to obtain the best set of parameters ($G_V$, $\alpha$, $\rho_{sc}$) that satisfies all the observational data up to date.

We arrange the present work as follows. In the next Sec. \ref{sec:NJL_Model}, we briefly discuss the salient features of the NJL model for SQM. We then depict the mechanism of invoking the interaction between DM and SQM via contact type of four-Fermi interactions in Sec. \ref{sec:DNJL_Model}. We also discuss the methodology for estimating the structural properties of DMSQSs in Sec. \ref{Sec:Structure}). The approach for Bayesian analysis of the free parameters of the DNJL model is discussed in Sec. \ref{Sec:Bayesian analysis}. The procedure for obtaining the $f$-mode oscillation frequency of the DMSQSs is demonstrated in Sec. \ref{Sec:oscillation}. We then present the results and corresponding discussions in Sec. \ref{Results}. We summarize and conclude in the final section, Sec. \ref{Conclusion} of the paper.

\section{NJL model for quark star}
\label{sec:NJL_Model}
In order to account for the the 3-flavor SQM (consisting of the $u$, $d$, and $s$ quarks along with the electrons), we consider the NJL model based on the effective QCD theory \cite{Nambu:1961tp}. Apart from the scalar, pseudo-scalar, and vector types of interactions, this model also includes the t’Hooft six-fermion interaction, which is required to break the axial symmetry \cite{Hatsuda:1987pc,Vogl:1991qt,Klevansky:1992qe,Hatsuda:1994pi,Kitazawa:2002jop,Lopes:2020rqn,Masuda:2012ed,Buballa:2003qv}. The corresponding Lagrangian is given as
\begin{eqnarray}
{\cal L}_{NJL} &=& \bar{\psi}_f[\gamma^\mu(i \partial_\mu - m_f)]{\psi}_f + G_S \sum_{a=0}^8\left[(\bar{\psi}_f \lambda_a \psi)^2 + (\bar{\psi}_f \gamma_5 \lambda_a \psi)^2\right] - G_V (\bar{\psi}_f \gamma^\mu \psi)^2 \nonumber \\ &-& K\left\lbrace det[\bar{\psi}(1+\gamma_5)\psi] + det[\bar{\psi}(1-\gamma_5)\psi]\right\rbrace.
\label{Eq:njl_lagrangian} 
\end{eqnarray}
where, $m_f$ are the current quark masses, $\lambda_a$ are the eight Gell-Mann flavor matrices and $G_S$, $G_V$ and $K$ are the coupling constants corresponding to the scalar channel, vector channel, and the t’Hooft interaction term, respectively. The vector interaction gives a universal repulsion among the three different flavors of the quarks. Unlike the relativistic models, there is no mediator in this model, and the quark-quark interaction is of a direct four-fermi contact type. Due to such a type of interaction, the scalar quark condensate of flavor $f$ is given as
\begin{eqnarray}
\phi_f=<\bar{\psi}\psi>=\frac{-\gamma_q M_f}{2\pi^2}\int_{{k_F}_f}^\Lambda \frac{k^2 dk}{\sqrt{M_f^2 + k^2}},
\label{eq:phi_f}
\end{eqnarray}
where, ($i$, $j$, $k$) corresponds to the cyclic
permutation of the $u$, $d$, and $s$ quarks. $M_f$, being the dressed mass of the quarks of flavor $f$ in the mean-field approximation, are generated dynamically through the NJL interactions. It is given as
\begin{eqnarray}
M_f=m_f - 4G_S \phi_f + 2K \phi_i \phi_j,~~~~~~   i\neq j\neq f.
\label{eq:M_f}
\end{eqnarray}
The color-spin degeneracy factor is $\gamma_q$=6 for the quarks. $\Lambda$ is the 3-momentum cut-off. The vector channel introduces displacement of the energy eigenvalues ($E_f$), which is equal to the chemical potential of the quarks ($\mu_f$) at $T$=0, and is given as
\begin{eqnarray}
E_f=\mu_f = \sqrt{M_f^2 + {k_F}_f^2} + 2G_V \rho_q,
\label{eq:E_f}
\end{eqnarray}
where, the total quark number density is expressed as the sum of the number density of each flavor as
\begin{eqnarray}
\rho_q=\sum_f \rho_f=\sum_f \frac{\gamma_q}{6\pi^2} {k_F}_f^3,
\label{eq:rho_q}
\end{eqnarray}
${k_F}_f$ being the Fermi momentum of the quarks of flavor $f$. The baryon density can be calculated as
\begin{eqnarray}
\rho=\frac{1}{3} \rho_q.
\label{eq:rho_B}
\end{eqnarray}
The EoS for SQM derived from the Lagrangian (Eq. (\ref{Eq:njl_lagrangian})) can be found in \cite{Lopes:2020rqn,Masuda:2012ed}. The energy density of the system is given as
\begin{eqnarray}
\varepsilon_{NJL}=\sum_f \Bigg[\frac{-\gamma_q}{2\pi^2}\int_{{k_F}_f}^\Lambda \sqrt{M_f^2 + k^2} ~k^2 dk + 2G_S \phi_f^2 \Bigg] + G_V \rho_q^2 - 4K \phi_i \phi_j \phi_k - \varepsilon_{vac} + \frac{\gamma_e}{2\pi^2}\int_0^{k_e} \sqrt{m_e^2 + k^2}~k^2 dk,
\label{Eq:njl_e}
\end{eqnarray}
where, $\gamma_{q} = \gamma_e$=2 and $\varepsilon_{vac}$ is the constant vacuum energy introduced to set the energy density of the physical vacuum ($k_F$=0) equal to zero. The pressure of the system is given as
\begin{eqnarray}
P_{NJL}=\sum_{i=u,d,s,e} \mu_i \rho_i - \varepsilon_{NJL}.
\label{Eq:njl_P}
\end{eqnarray}
The charge neutrality and the chemical equilibrium conditions are, respectively, imposed as
\begin{eqnarray}
\sum_{i=u,d,s,e} q_i \rho_i = 0
\label{eq:charge_neutrality} 
\end{eqnarray}
and 
\begin{eqnarray}
\mu_d=\mu_s=\mu_u+\mu_e.
\label{eq:chemical_equilibrium} 
\end{eqnarray}
\subsubsection*{NJL model parameters}
\label{sec:NJL parameters}
We consider the Hatsuda-Kunihiro (HK) parameter set from \cite{Hatsuda:1994pi} for the NJL model for SQM as shown in Tab. \ref{Tab:NJL}.
\begin{table}[!ht]
\caption{Parameter set of the NJL model stating the current mass of the three quarks ($m_{u,d,s}$), the momentum cut-off ($\Lambda$), the $\Lambda$ scaled couplings $G_S$ and $K$.}
{{
\setlength{\tabcolsep}{33pt}
\begin{center}
\begin{tabular}{ c c c c c c c c}
\hline
\hline
$m_{u,d}$ (MeV) & $m_s$ (MeV) & $\Lambda$ (MeV) & $G_S \Lambda^2$ & $K \Lambda^5$ \\
\hline
5.5 & 135.7 & 631.4 & 3.67 & 9.29 \\
\hline
\hline
\end{tabular}
\end{center}
}}
\protect\label{Tab:NJL}
\end{table}  
The value of $G_V$ is not very well known in the literature. In order to reproduce the LQCD results, the authors have considered 0.2$G_S<G_V<$ 0.3$G_S$ to study the interplay between chiral transition and color-superconducting phase \cite{Kitazawa:2002jop}. In order to reproduce the slope of the pseudo-critical temperature for the chiral phase transition at low chemical potential extracted from LQCD simulations, the range was considered to be 0.283$G_S <G_V<$ 0.373$G_S$ \cite{Contrera:2012wj}, 0.25$G_S<G_V<$ 0.4$G_S$ \cite{Kashiwa:2011td}, and $G_V$=0.33$G_S$ \cite{Sugano:2014pxa,Lopes:2020rqn}. In \cite{Menezes:2014aka,Hanauske:2001nc} $G_V$=$G_S$ has been considered while studies of the Polyakov-NJL model applied to the QCD phase diagram suggest that $G_V$ may be larger than $G_S$ \cite{Bratovic:2012qs,Lourenco:2012yv} and therefore \cite{Masuda:2012ed} have considered up to $G_V$=1.5$G_S$. In the present work, we consider two values of the ratio $G_V/G_S$. One value is taken to be 0.3 to be consistent with that suggested by LQCD. Another value is considered to be 0.5, which is a little higher than that suggested by LQCD.

\section{Dark matter admixed quark star model}
\label{sec:DNJL_Model}
We introduce the interaction between the quarks and the DM fermion ($\chi$) via the four-fermi contact type. This is equivalent to the heavy mediator scenario. Similar to the SQM, we introduce a scalar and a vector channel in the dark sector. The complete Lagrangian for the DNJL model is given as
\begin{eqnarray}
{\cal L}_{DNJL} &=& {\cal L}_{NJL} + \bar{\chi}[\gamma^\mu(i \partial_\mu - m_{\chi})]{\chi} + G_{SD}[(\bar{\chi} \chi)^2 + (\bar{\chi} \gamma_5 \chi)^2] -G_{VD}(\bar{\chi} \gamma^{\mu} \chi)(\bar{\chi} \gamma_{\mu} \chi) \nonumber\\
&+&G_{SqD}[(\bar{\chi} \chi)(\bar{\psi}_f {\psi}_f) + (\bar{\chi} \gamma_5 \chi)(\bar{\psi}_f \gamma^5 {\psi}_f)] - G_{VqD}(\bar{\chi} \gamma^{\mu} \chi)(\bar{\psi}_f \gamma_{\mu} {\psi}_f) .
\protect\label{Eq:Dnjl_lagrangian}
\end{eqnarray}
In Eq. (\ref{Eq:Dnjl_lagrangian}), the terms with couplings $G_{SD}$ and $G_{VD}$ denote the DM self-interaction via the scalar and vector channels, respectively. On the other hand, the DM-quark interaction is invoked via the scalar and vector channels, as depicted by the terms with couplings $G_{SqD}$ and $G_{VqD}$, respectively. We assume that the DM-quark couplings are same for $u$, $d$, and $s$ quarks, i.e., $G_{SuD}=G_{SdD}=G_{SsD}$ and $G_{VuD}=G_{VdD}=G_{VsD}$. The DM self-interaction couplings are considered in terms of the DM momentum cut-off as
\begin{eqnarray}
G_{SD}=\frac{1}{\Lambda_{SD}^2}~~;~~ G_{VD}=\frac{1}{\Lambda_{VD}^2},
\end{eqnarray}
and the DM-quark interaction couplings are taken in terms of the EFT expansion scale for the quark-DM interaction as
\begin{eqnarray}
G_{SqD}=\frac{1}{\Lambda_{SqD}^2}~~;~~ G_{VqD}=\frac{1}{\Lambda_{VqD}^2}.
\end{eqnarray}
The scalar-type of DM-quark interaction modifies the dressed mass of the quark, and therefore Eq. (\ref{eq:M_f}) is modified as
\begin{eqnarray}
M_f=m_f - 4G_S \phi_f + 2K \phi_i \phi_j - 2G_{SqD} \phi_{\chi},
\label{eq:D-M_f}
\end{eqnarray}
while the vector-type of DM-quark interaction contributes to the chemical potential of the quarks. Therefore, Eq. (\ref{eq:E_f}) is now modified as
\begin{eqnarray}
E_f=\mu_f = \sqrt{M_f^2 + {k_F}_f^2} + 2G_V \rho_q + G_{VqD} \rho_{\chi}.
\label{eq:D-E_f}
\end{eqnarray}
In Eq. (\ref{eq:D-M_f}), $\phi_{\chi}$ is the scalar condensate of the fermionic DM given by
\begin{eqnarray}
\phi_{\chi}=<\bar{\chi}\chi>=\frac{-\gamma_{\chi} M_{\chi}}{2\pi^2}\int_{{k_F}_{\chi}}^{\Lambda_{SD}} \frac{k_{\chi}^2 dk_{\chi}}{\sqrt{M_{\chi}^2 + k_{\chi}^2}},
\label{eq:phi_chi}
\end{eqnarray}
where, $\gamma_{\chi}$=2 and $M_{\chi}$ is the dressed mass of the fermionic DM. It is given as
\begin{eqnarray}
M_{\chi}=m_{\chi} - 4G_{SD} \phi_{\chi} + 2G_{SqD}\sum_f \phi_f.
\label{eq:M_chi}
\end{eqnarray}
In Eq. (\ref{eq:D-E_f}), $\rho_{\chi}$ is the number density of the fermionic DM. In the present work, we consider the relation between the DM and the baryon density as
\begin{eqnarray}
\rho_{\chi}= \rho_{sc} \alpha(e^{\rho/\rho_{sc}} - 1),
\label{Eq:rho_chi}
\end{eqnarray}
where, $\rho_{sc}$ is the reference number density. We initially take $\rho_{sc}$=1 fm$^{-3}$ in order to keep $\alpha$ as the only free parameter that controls the behavior and distribution of DM relative to the baryonic matter within the star. We first aim to constrain the value of $\alpha$ in light of the different bounds on the DM parameters and the astrophysical constraints on the structural properties of compact stars. The reason for considering the density profile of the DM particularly in the form of Eq. (\ref{Eq:rho_chi}) is that it includes the contributions from the different powers of $\rho$. In this way the DM Fermi momentum ${k_F}_{\chi}$ is no longer constant throughout the density range of the star. Thereby, we avoid the oversimplification of the DM dynamics at high density and our present consideration (Eq. (\ref{Eq:rho_chi})) is also on par with the fact that gravitational effects lead to a variable DM density profile in DMSQSs as suggested by \cite{Kumar:2025ytm}. Moreover, since ${k_F}_{\chi}$ is variable with the density (radius) of the star, it does not introduce any thermodynamic inconsistency \cite{Hajkarim_2025}. We later vary $\rho_{sc}$ along with $\alpha$ to perform Bayesian analysis in order to set constraints on them from the various astrophysical observational data. 

The DM Fermi momentum ${k_F}_{\chi}$ is then written as
\begin{eqnarray}
{k_F}_{\chi}={\left(\frac{6\pi^2}{\gamma_{\chi}}\rho_{\chi}\right)}^{1/3}.
\end{eqnarray}
The chemical potential of the DM is also expressed as
\begin{eqnarray}
E_{\chi}=\mu_{\chi} = \sqrt{M_{\chi}^2 + {k_F}_{\chi}^2} + 2G_{VD} \rho_{\chi} + G_{VqD} \rho_q.
\label{eq:E_chi}
\end{eqnarray}
It can be noticed from Eqs. (\ref{eq:D-M_f}), (\ref{eq:D-E_f}), (\ref{eq:M_chi}), and (\ref{eq:E_chi}) that both quarks and DM contribute to the dressed mass and the chemical potential of the quarks and DM due to the DM-quark interaction.

The energy density of the DMASQM is given as
\begin{eqnarray}
\varepsilon_{DNJL}=\varepsilon_{NJL} + \frac{-\gamma_{\chi}}{2\pi^2}\int_{{k_F}_{\chi}}^{{\Lambda}_{SD}} \sqrt{M_{\chi}^2 + k_{\chi}^2} ~k_{\chi}^2 dk_{\chi} + 2G_{SD} \phi_{\chi}^2 + 
G_{VD} \rho_{\chi}^2 + 2G_{SqD} \phi_f \phi_{\chi} + G_{VqD} \rho_q \rho_{\chi} - {\varepsilon_{vac}}^{\prime},
\label{Eq:Dnjl_e}
\end{eqnarray}
where, ${\varepsilon_{vac}}^{\prime}$ is the modified vacuum energy due to the DM-quark interaction. The pressure of the DMASQM is now given by
\begin{eqnarray}
P_{DNJL}=\sum_{i=u,d,s,e,\chi} \mu_i \rho_i - \varepsilon_{DNJL}.
\label{Eq:Dnjl_P}
\end{eqnarray}
Since we include charge neutral DM, the DMASQM remains charge neutral following Eq. (\ref{eq:charge_neutrality}).

\subsubsection*{Dark matter model parameters}
\label{sec:DM parameters}

It is very crucial to choose the parameters of the dark sector carefully in case of contact interaction. We determined the parameters for the dark sector from the constraints of DM self-interaction and DM relic density bounds, consistent with the corresponding observations from direct detection experiments, LZ, XENON, DarkSide, CRESST, and collider experiment LHC. DM pair annihilation cross section can be theoretically obtained in terms of $\Lambda_{SqD}$ and $\Lambda_{VqD}$ using Eq. (\ref{Eq:Dnjl_lagrangian}). Then we calculate the values of $\Lambda_{SqD}$ and $\Lambda_{VqD}$ for different values of DM mass ($m_\chi$) using MicrOMEGAs~\cite{Belanger:2013oya} which satisfy the present-day non-baryonic relic abundance. In this scenario, the quark-antiquark pair production happens because of the dominant thermal annihilation channel of the DM-DM pair. The calculated values of $\Lambda_{SqD}$ and $\Lambda_{VqD}$ have been listed in Tab. \ref{Tab:DNJL}.

The idea of self-interactions of DM can solve the riddles of the core-cusp problem, missing satellites problem, and too-big-to-fail problem. To understand the magnitude of DM self-interactions and their effect in the small scale structures, N-body simulations have been the primary tools over the last few decades. The fundamental assumption in the majority of these simulations is energy-independent and isotropic scattering with a contact-type interaction~\cite{Tulin:2013teo, Tulin:2017ara}. DM self-interaction is quantified by $\sigma/m_\chi$, where $\sigma$ is the self-interaction cross section. A wide range of this quantity has been scanned by these simulations ($\sigma/m_\chi \approx 0.1 $-$ 50~\rm{cm^2/gm}$). In dwarf galaxy scales, DM self-interaction is well described by $\sigma/m_\chi \sim 0.1 $-$ 10~\rm{cm^2/gm}$, whereas, to solve the core-cusp problem and too-big-to-fail problem, several studies have converged on the favorable range $\sigma/m_\chi \sim 0.5 $-$ 1~\rm{cm^2/gm}$. On the other hand, in large galaxy scales, e.g, Milky Way, and in cluster scales, e.g, Bullet cluster, the requirement for the DM self-interaction is $\sigma/m_\chi \sim 0.1 $-$ 1~\rm{cm^2/gm}$~\cite{Tulin:2013teo, Tulin:2017ara}. In this work, in order to be consistent with the astrophysical observations on various scales, we work with a conservative value of the DM self-interaction $\sigma/m_\chi = 1~\rm{cm^2/gm}$ and equate the theoretically obtained expression of $\sigma/m_\chi$ (using Eq. (\ref{Eq:Dnjl_lagrangian})) to it to extract the values of $\Lambda_{SD}$ and $\Lambda_{VD}$ as a function of DM mass $m_\chi$. The calculated values of $\Lambda_{SD}$ and $\Lambda_{VD}$ have been listed in Tab. \ref{Tab:DNJL}.
\begin{table}[!ht]
\caption{SGP parameterization of the DM model stating the DM mass ($m_{\chi}$), and the corresponding DM momentum cut-offs $\Lambda_{SD}$, $\Lambda_{VD}$, $\Lambda_{SqD}$, and $\Lambda_{VqD}$.}
{{
\setlength{\tabcolsep}{25pt}
\begin{center}
\begin{tabular}{ c c c c c c c c}
\hline
\hline
$m_{\chi}$ (MeV) & $\Lambda_{SD}$ (GeV) & $\Lambda_{VD}$ (GeV) & $\Lambda_{SqD}^{max}$ (GeV) & $\Lambda_{VqD}^{max}$ (GeV) \\
\hline
5 $\times$ 10$^3$ & 0.108 & 0.134 & 295.0 & 345.0 \\
100 & 0.041 & 0.050 & 3.06 & 4.0  \\
125 & 0.043 & 0.053 & 24.0 & 30.8 \\
150 & 0.045 & 0.056 & 42.0 & 50.5 \\
\hline
\hline
\end{tabular}
\end{center}
}}
\protect\label{Tab:DNJL}
\end{table}  

\section{Structural properties of dark matter admixed quark stars}
\label{Sec:Structure}
Using the EoS of DMASQM, we compute the global properties such as the gravitational mass ($M$) and the radius ($R$) of the DMSQSs.  For a static spherically symmetric star, the metric is given as
\begin{eqnarray}
ds^2=-e^{2\Phi(r)}dt^2 + e^{2\lambda(r)}dr^2 + r^2d\theta^2 + r^2 \sin^2\theta d\phi^2
\label{metric}
\end{eqnarray}
where, $\Phi$ and $\lambda$ are the metric functions. The Einstein field equations are solved for the given metric for an ideal fluid to obtain the well-known Tolman-Oppenheimer-Volkoff (TOV) equations \cite{Tolman:1939jz,Oppenheimer:1939ne} and are given as 
\begin{eqnarray}
\frac{dP(r)}{dr}&=&-\Big(\varepsilon(r)+P(r)\Big)\frac{d\Phi(r)}{dr},
\label{eq:tov}\\
\frac{d\Phi(r)}{dr}&=&\frac{M(r)+4\pi r^3 P(r)}{r\Big(r-2 M(r)\Big)},
\label{eq:tov2}\\
\frac{dM(r)}{dr}&=& 4\pi r^2 \varepsilon(r).
\label{eq:tov3}
\end{eqnarray} 
The coupled differential equations are integrated from the center ($r=0$) to the surface ($r=R$) of the star. By solving Eqs. (\ref{eq:tov}), (\ref{eq:tov2}), and (\ref{eq:tov3}) for all possible values of central energy density $\varepsilon_c$, we obtain the mass $M$ and the radius $R$ of the star. The mass function $M(r)=r(1-e^{-2\lambda(r)})/2$ satisfies Eq. (\ref{eq:tov3}). The global quantity, compactness, defined as $C=M/R$, quantifies stellar gravity.
 
The tidal deformability is the quadrupole deformation of a star in a binary system due to the tidal field of its companion star. For an isolated star, the tidal deformability is calculated in the limit where the source of the static external quadrupolar tidal field is very far away \cite{Hinderer:2007mb}. We obtain the dimensionless tidal deformability ($\Lambda$) by following the \cite{Hinderer:2007mb,Hinderer:2009ca} and is given as
\begin{eqnarray} 
\Lambda=\frac{2}{3} k_2 R^5.
\label{Eq:Lambda}
\end{eqnarray}
Here $k_2$ is the tidal love number, which is given in terms of the compactness ($C$), and a quantity ($y$) \cite{Hinderer:2007mb,Hinderer:2009ca}. In the case of QSs, $y$ is defined as \cite{Hinderer:2009ca,Kumar:2022byc}
\begin{eqnarray}
y=\frac{RH'(R)}{H(R)} - \frac{4\pi R^3 \varepsilon_s}{M(R)};
\label{y}
\end{eqnarray}
where, $\varepsilon_s$ is the energy density at the surface of the QS.

\section{Bayesian analysis of the parameters $G_V/G_S$ and $\alpha$ with respect to the observational constraints}
\label{Sec:Bayesian analysis}

The free parameters of the DNJL model can be constrained in the light of the existing observations regarding the structural properties of compact stars, e.g., mass, radius, and tidal deformability. For this purpose we take into account the following data sets listed below in Tab. \ref{Tab:constraints}:
\begin{table}[!ht]
\caption{The constraints on the structural properties of compact stars used for performing Bayesian analysis.}
{{
\setlength{\tabcolsep}{25pt}
\begin{center}
\begin{tabular}{ c c c c c c c c}
\hline
\hline
Source & Constraint & Refs. \\
\hline
PSR J0030+0451 & $M = 1.44^{+0.15}_{-0.14}~M_\odot$ and $R = 13.02^{+1.24}_{-1.06}~\rm{km}$ & Miller et al.~\cite{Miller:2019cac} \\ \\
PSR J0740+6620 & $M = 2.08^{+0.07}_{-0.07}~M_\odot$ and $R = 12.92^{+2.09}_{-1.13}~\rm{km}$ & Dittmann et al.~\cite{Miller:2021qha} \\ \\
PSR J0437-4715 & $M = 1.418^{+0.037}_{-0.037}~M_\odot$ and $R = 11.36^{+0.95}_{-0.63}~\rm{km}$ & Choudhury et al.~\cite{Choudhury:2024xbk} \\ \\
PSR J1231-1411 & $M = 1.04^{+0.05}_{-0.03}~M_\odot$ and $R = 12.6^{+0.3}_{-0.3}~\rm{km}$ & Salmi et al.~\cite{Salmi:2024bss} \\ \\
PSR J0614-3329 & $M = 1.44^{+0.06}_{-0.07}~M_\odot$ and $R = 10.29^{+1.01}_{-0.86}~\rm{km}$ & Mauviard et al.~\cite{Mauviard:2025dmd} \\ \\
HESS J1731-347 & $M = 0.77^{+0.20}_{-0.17}~M_\odot$ and $R = 10.04^{+0.86}_{-0.78}~\rm{km}$ & Doroshenko et al.~\cite{Doroshenko:2022} \\ \\
GW170817 & $\Lambda_{1.4} = 190^{+390}_{-120}$ & Abbott et al.~\citep{LIGOScientific:2018cki} \\
\hline
\hline
\end{tabular}
\end{center}
}}
\protect\label{Tab:constraints}
\end{table}  
We infer the posterior distribution of the free parameters ($\alpha, G_V/G_S, \rho_{sc}$), defining the EoS, in light of the observational data listed above. Consequently, we constrain the EoS to satisfy the observations. Bayes' theorem describes the posterior distribution of the parameters ($\alpha, G_V/G_S, \rho_{sc}$) for given observational data $\mathcal{D}$ as
\begin{eqnarray}
p\left((\alpha, G_V/G_S, \rho_{sc}) \vert \mathcal{D} \right) = \frac{p\left( \mathcal{D} \vert (\alpha, G_V/G_S, \rho_{sc})  \right) p(\alpha, G_V/G_S, \rho_{sc})}{p(\mathcal{D})},
\end{eqnarray} 
where, $p\left( \mathcal{D} \vert (\alpha, G_V/G_S, \rho_{sc}) \right)$ is the full likelihood in terms of the given observational data, $p(\alpha, G_V/G_S, \rho_{sc})$ is the prior distribution of the parameters and $p(\mathcal{D})$ is the evidence. We follow~\cite{Ayriyan:2024zfw} to estimate the full likelihood and the prior distribution is considered to be equiprobable within a range of reasonable guess values obtained from initial rough estimation. So,
\begin{eqnarray}
p(\alpha, G_V/G_S, \rho_{sc}) = \frac{1}{\bigg\lvert \left(\alpha, G_V/G_S, \rho_{sc}\right)\bigg\rvert} = \frac{1}{N}.
\end{eqnarray}
We calculate the likelihoods for all the independent observations ($\mathcal{D}_{j}$) and then the full likelihood is obtained by the overall product as
\begin{eqnarray}
p\left( \mathcal{D} \vert (\alpha, G_V/G_S, \rho_{sc})  \right) = \prod_{j} p\left( \mathcal{D}_{j} \vert (\alpha, G_V/G_S, \rho_{sc})  \right).
\end{eqnarray}
The evidence is estimated by summing all possible values of ($G_V/G_S$, $\alpha$, and $\rho_{sc}$) taken into consideration, 
\begin{eqnarray}
p(\mathcal{D}) = \sum_{(\alpha, G_V/G_S, \rho_{sc})} p\left( \mathcal{D} \vert (\alpha, G_V/G_S, \rho_{sc})  \right) p(\alpha, G_V/G_S, \rho_{sc}).
\end{eqnarray}
We compute the likelihood for each of the mass-radius constraints related to the observational pulsar data by integrating over the central energy density $\varepsilon_c$ with the appropriate probability density function~\cite{Ayriyan:2024zfw},
\begin{eqnarray}
p\left( \mathcal{D}_{MR^{(i)}} \vert (\alpha, G_V/G_S, \rho_{sc})  \right) = \int_{\varepsilon_c^{min}}^{\varepsilon_c^{max}(\alpha, G_V/G_S, \rho_{sc})} f_{MR}^{(i)}  \left( M(\epsilon_c ; \alpha, G_V/G_S, \rho_{sc}), R(M) \right) ~pr(\alpha, G_V/G_S, \rho_{sc})~d \epsilon_c ,
\end{eqnarray}
where, $ pr(\alpha, G_V/G_S, \rho_{sc}) = 1/(\varepsilon_c^{max}(\alpha, G_V/G_S, \rho_{sc}) - \varepsilon_c^{min})$. Following~\cite{Ayriyan:2024zfw}, we construct the probability density functions $f_{MR}^{(i)}$ using Kernel Density Estimation (KDE)~\cite{ChaconDuong:2018} on the basis of the data acquired from the Zenodo repository for pulsars (M-R constraints): PSR J0030+0451~\cite{miller_2019_3473466}, PSR J0740+6620~\cite{dittmann_2024_10215109}, PSR J0437-4715~\cite{choudhury_2024_13766753}, PSR J1231-1411~\cite{salmi_2024_13358349}, PSR J0614-3329~\cite{mauviard_2025_15603406}, and HESS J1731-347~\cite{doroshenko_2022_8232233}. For GW data, the likelihood is estimated in a similar fashion as an integration over $\varepsilon_c$,
\begin{eqnarray}
p\left( \mathcal{D}_{GW} \vert (\alpha, G_V/G_S, \rho_{sc})  \right) = \int_{\varepsilon_c^{min}}^{\varepsilon_c^{max}(\alpha, G_V/G_S, \rho_{sc})} f_{GW} \left( \Lambda_1(\epsilon_c ; \alpha, G_V/G_S, \rho_{sc}), \Lambda_2(\Lambda_1) \right) ~pr(\alpha, G_V/G_S, \rho_{sc})~d \epsilon_c,
\end{eqnarray}
where, the probability density function $f_{GW}$ is built using the KDE based on the GW170817 data~\cite{ligoLIGOP1800115v12GW170817}.

\section{Non-radial oscillation properties of dark matter admixed quark stars}
\label{Sec:oscillation}

We calculate the non-radial oscillation frequency of the DMSQSs by adopting both the formalisms of full general relativistic consideration and also using the Cowling approximations, thereby neglecting the metric perturbations. For the first approach, we follow \cite{1967ApJ...149..591T,Lindblom:1983ps,Lu:2011zzd,Guha:2025ssq,Thakur:2024ejl} while for the second one we follow \cite{Sotani:2010mx}. The details of the two formalisms are discussed in Appendix \ref{Sec:Appendix}. The Cowling approximation and GR approach are depicted in Sec. \ref{sec:Cowling} and \ref{sec:GR}, respectively.

\subsubsection*{Observational prospect}
\label{sec:Observational prospect}

The amplitude of the GW strain depends on the oscillation amplitudes. In general, the $f$-mode oscillation can be used to extract energy at the level of oscillations and that drives the emission of GWs, thereby $E_{GW}=E_{osc}$. The lowest order post-Newtonian quadrupole
formula for the GW radiation power $P_{GW}$ connects the oscillation energy with the GW damping time $\tau$ as $\tau=\frac{2E_{GW}}{P_{GW}}$ \cite{Thorne:1969rba}. The approximation is quite accurate, with maximum relative deviations of only 7\% compared to the full solutions for cold compact stars \cite{Zheng:2024tjl, Zheng:2025xlr}. The amplitude of the GW strain $h_+$ depends on $P_{GW}$ and is given as \cite{Zheng:2024tjl, Zheng:2025xlr}
\begin{eqnarray}
h_+=\frac{{\sqrt{30E_{GW}/\tau}}}{2\pi D f}
\label{eq:strain}
\end{eqnarray}
where, $D$ is the distance from the source and $\tau$ is the damping time obtained from the full GR \cite{Zheng:2024tjl, Zheng:2025xlr}. One of the prominent ways in which the $f$-mode may be triggered is the glitching behavior of pulsars \cite{Ho:2020nhi,Haskell:2023exo}. Assuming that the entire glitch energy is emitted as GW energy, the GW strain may be calculated corresponding to the $f$-mode oscillation using Eq. (\ref{eq:strain}). The estimated energies of observed pulsar glitches may vary in the range $10^{37}-10^{44}$ ergs, and the majority of the known glitching pulsars are at distances $D < 6$ kpc \cite{Ho:2020nhi}. Using this information, we can estimate the GW strain amplitude obtained with Eq. (\ref{eq:strain}) of GW radiation emitted via $f$-mode oscillation of DMSQSs. The estimation can be compared with the projected sensitivity of upcoming GW detectors like the aLIGO, A+, CE1, and ET. 

\section{Results}
\label{Results}
\subsection{Dark matter admixed quark matter}
\label{subsec:DMQM}
\subsubsection*{Choice of benchmark values of $m_{\chi}$ and corresponding values of $\Lambda_{SqD}$ and $\Lambda_{VqD}$}
\begin{figure*}[ht]
\centering
\subfigure[\ For scalar-type interaction]{\includegraphics[width=0.48\linewidth,height=7cm]{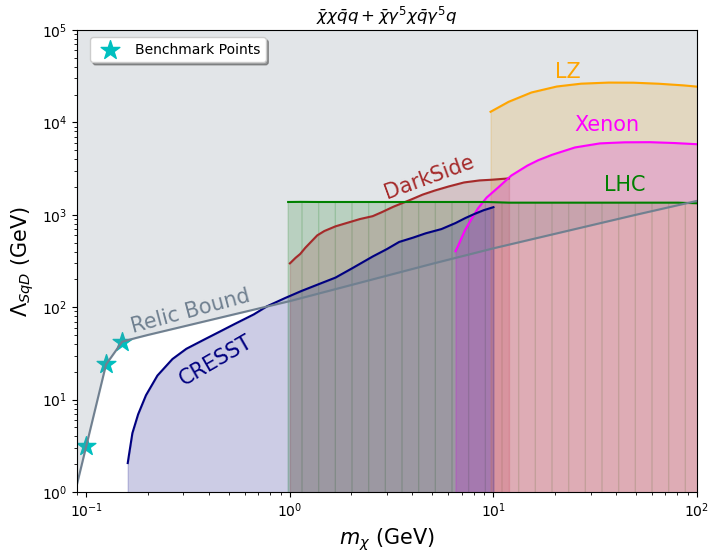} \label{fig:mchi_LamSqD}}
\hfill 
\subfigure[\ For vector-type interaction]{\includegraphics[width=0.48\linewidth,height=7cm]{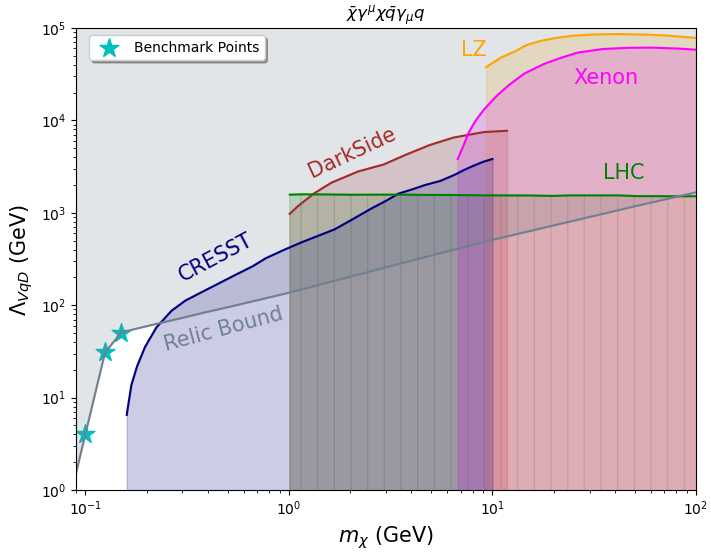}\label{fig:mchi_LamVqD}}
\caption{Variation of EFT expansion scale for quark-DM interaction with mass of dark matter. The shaded regions are ruled out from direct detection experiments, LZ \cite{LZ:2022lsv}, XENON \cite{XENON:2019rxp}, DarkSide \cite{DarkSide:2018bpj}, CRESST \cite{CRESST:2019jnq} and collider experiment LHC \cite{Roy:2024ear}. For the present work we choose the benchmark points (marked with asterisks) from the allowed region.}
\label{fig:mchi_LamSqD_LamVqD}
\end{figure*}
The choice of $m_{\chi}$ and the corresponding values of the EFT expansion scale for quark-DM interaction $\Lambda_{SqD}$ and $\Lambda_{VqD}$ are discussed in Sec. \ref{sec:DM parameters}. In Fig. \ref{fig:mchi_LamSqD_LamVqD} we present the variation of $\Lambda_{SqD}$ and $\Lambda_{VqD}$ with $m_{\chi}$. The scalar type of interaction is shown in Fig. \ref{fig:mchi_LamSqD} while the vector type in Fig. \ref{fig:mchi_LamVqD}. It can be seen that in the $m_{\chi}-\Lambda_{SqD}$ and $m_{\chi}-\Lambda_{VqD}$ planes, there is a very narrow space indicating the allowed region (non-shaded region). Almost the entire parameter spaces for both scalar and vector types of interaction are excluded (shaded regions) by various direct and indirect detection experiments (LZ, XENON, DarkSide, CREST), collider (LHC), and the relic bound. The allowed region falls in the light DM range $\sim$MeV. Within the narrow allowed region, we choose three benchmark points as $m_{\chi}$=100, 125, and 150 MeV with corresponding values of $\Lambda_{SqD}$ and $\Lambda_{VqD}$ for the present work. After choosing the values of $m_{\chi}$ in the allowed sub-GeV regime, we obtain the corresponding values of the EFT expansion scale for DM self-interaction ($\Lambda_{SD}$ and $\Lambda_{VD}$). The values of $\Lambda_{SqD}$, $\Lambda_{VqD}$, $\Lambda_{SD}$, and $\Lambda_{VD}$ for each value of $m_{\chi}$ are already shown in Tab. \ref{Tab:DNJL} as the SGP parameterization.

With the parameters obtained from the dark sector, we perform our initial analysis of the properties of DMSQSs considering each of the two values of $G_V/G_S = 0.3, 0.5 $, and each of the three values of $m_{\chi}=100, 125, 150$ MeV, with $\alpha$ as 0.1, 0.2, and 0.3. For this purpose, we keep $\rho_{sc} = 1~\rm{fm^{-3}}$ at this stage. Later we perform Bayesian analysis to constrain the free parameters of the model, viz. $\alpha, G_V/G_S $ and $\rho_{sc}$, in the light of recent observations, as formulated in Sec. \ref{Sec:Bayesian analysis}.

\subsubsection*{Equation of state}

\begin{figure*}[t]
\centering
\subfigure[\ $m_{\chi}$=5 GeV]{\includegraphics[width=0.48\linewidth,height=7cm]{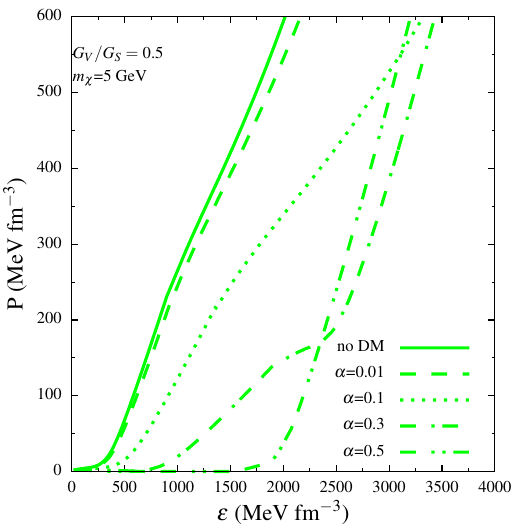}\label{fig:eos_Dnjl_mchi5G}}
\hfill 
\subfigure[\ $m_{\chi}$=100 MeV]{\includegraphics[width=0.48\linewidth,height=7cm]{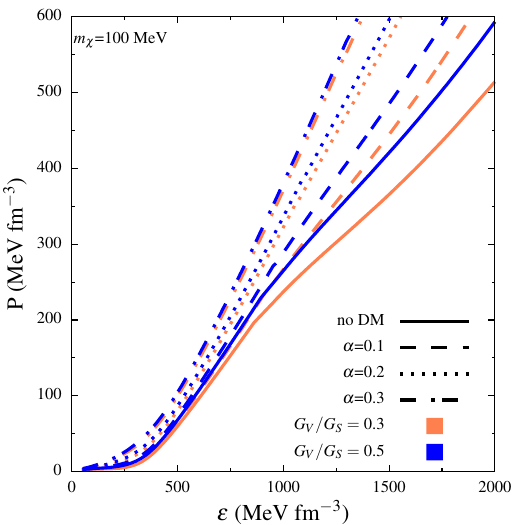}\label{fig:eos_Dnjl_mchi100M}}
\caption{Equation of state of dark matter admixed strange quark star for $G_V/G_S$=0.3, 0.5 and different values of $\alpha$ with (a) $m_{\chi}$=5 GeV and (b) $m_{\chi}$=100 MeV.}
\label{fig:eos_Dnjl}
\end{figure*}
With the initial choice of the parameters related to SQM and the dark sector, we compute the EoS of DMSQSs using Eqs. (\ref{Eq:Dnjl_e}) and (\ref{Eq:Dnjl_P}). We display the EoS of the DMSQSs in Fig. \ref{fig:eos_Dnjl}. It is already seen from \cite{Lopes:2020rqn} that the EoS of the SQSs (the no-DM case) is too soft for $G_V/G_S$=0.3 and 0.5. Therefore, it does not satisfy the maximum mass constraint for compact stars from PSR J0740+6620 \cite{Fonseca:2021wxt}. Therefore, we require the EoS of DMSQSs to be stiffer than that of the SQSs. From Fig. \ref{fig:eos_Dnjl_mchi5G} we find that for $m_{\chi}$=5 GeV, the EoS of DMSQS with $G_V/G_S$=0.5 is softer than that of the EoS of SQS for different values of $\alpha$. The DMSQS EoS will soften further if we consider $G_V/G_S$=0.3 because the EoS softens with a weaker repulsive interaction between the quarks. Therefore, the EoS of DMSQS with $m_{\chi}$=5 GeV will not satisfy the maximum mass constraint for compact stars from PSR J0740+6620. This implies that for the contact-type interaction between DM and SQM, heavy DM ($m_{\chi}>$ 1 GeV) is not only ruled out by LHC but also from the massive pulsar observations. Therefore, we proceed with the benchmark values of $m_{\chi}$ in the sub-GeV scale. With the first benchmark point ($m_{\chi}$=100 MeV) chosen from Fig. \ref{fig:mchi_LamSqD_LamVqD}, we show the EoS of DMSQS with $G_V/G_S$=0.3 and 0.5 in Fig. \ref{fig:eos_Dnjl_mchi100M}. For all the values of $\alpha$, the EoS of the DMSQS stiffens compared to the no-DM case. The stiffening is enhanced by larger values of $\alpha$. Therefore, it is expected that the DMSQSs will be more massive than the SQSs without DM and the maximum mass of the DMSQSs will increase with increasing values of $\alpha$.
\begin{figure*}[ht]
\centering
\subfigure[\ Particle fraction]{\includegraphics[width=0.48\linewidth,height=7cm]{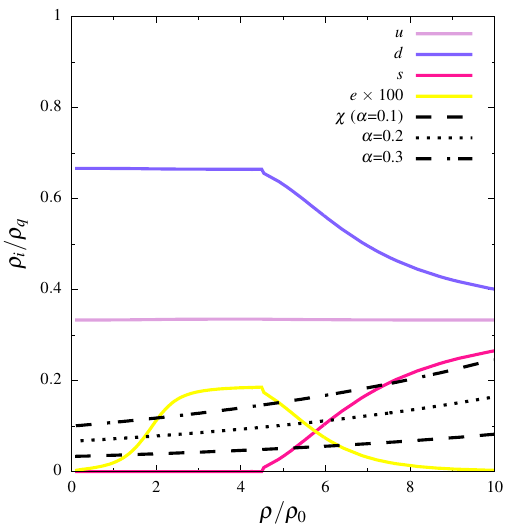} \label{fig:pf_mchi100M}}
\hfill 
\subfigure[\ Dressed mass]{\includegraphics[width=0.48\linewidth,height=7cm]{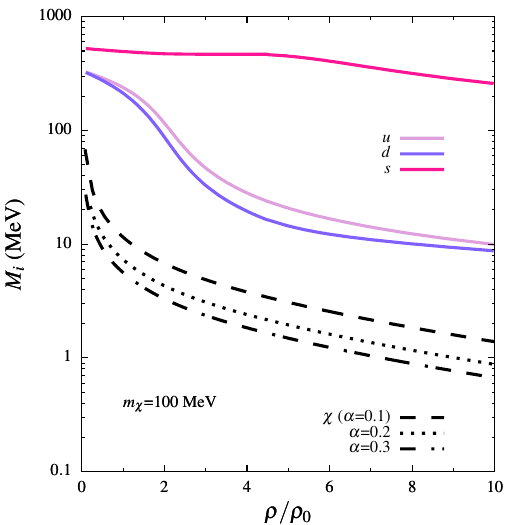} \label{fig:dressed_mass_Dnjl_mchi100M}}
\caption{Variation of (a) particle fraction and (b) dressed mass with baryon density of dark matter admixed strange quark star for different values of $\alpha$.}
\label{fig:DressedMass_pf100M}
\end{figure*}

Before proceeding to compute the structural properties of the DMSQSs, it is interesting to study the population of quarks and DM in the DMSQSs in Fig \ref{fig:pf_mchi100M}. Alongside, in Fig. \ref{fig:dressed_mass_Dnjl_mchi100M} we also depict the variation of the dressed mass of the quarks and DM in DMSQSs for $m_{\chi}$=100 MeV. As is known from \cite{Masuda:2012ed}, the particle fraction and dressed mass of the quarks do not depend on the value of $G_V/G_S$. In the DNJL model, the particle fraction and dressed mass of the quarks are feebly affected by DM following Eq. (\ref{eq:D-M_f}) since the value of $G_{SqD}$ is very small. Our values of $M_f$ and $\rho_{f,e}$ are consistent with those obtained by \cite{Masuda:2012ed}. Fig. \ref{fig:pf_mchi100M} shows that at low baryon density ($<$ 4.5 $\rho_0$), the system is composed of only the $u$ and $d$ quarks and the electrons. The combined effects of charge neutrality and $\beta$-equilibrium make $\rho_d$ twice that of $\rho_u$. The fraction of the $u$ quark remains constant throughout the baryon density profile. To maintain the charge neutrality condition, the fractions of the $d$ quark and the electron decrease as soon as the $s$ quark appears around 4.5 $\rho_0$. The value of $\rho_{\chi}$ (and therefore ${k_F}_{\chi}$) increases towards the core following Eq. (\ref{Eq:rho_chi}) and the fraction of DM increases with increasing values of $\alpha$. Fig. \ref{fig:dressed_mass_Dnjl_mchi100M} implies that before the appearance of the $s$ quark, $M_{u,d}$ decreases sharply because in this density regime the interaction among quarks is very strong via the $G_S$ term in Eq. (\ref{Eq:njl_lagrangian}). This leads to partial restoration of chiral symmetry. $M_s$ is slightly affected due to the coupling between the different flavors through the $K$ term in Eq. (\ref{Eq:njl_lagrangian}) \cite{Masuda:2012ed}. The different values of $m_{\chi}$ hardly affect the values of $M_f$ and $M_{\chi}$ is also very less affected by the presence of quarks since the value of $G_{SqD}$ in Eq. (\ref{eq:M_chi}) is very small. Like quarks, the value of $M_{\chi}$ also decreases from the rest mass of DM ($m_{\chi}$=100 MeV) along the density profile of the star. Since $M_{\chi}$ depends on $\rho_{\chi}$ (determined by $\alpha$), the value of $M_{\chi}$ decreases further with increasing values of $\alpha$.
\begin{figure*}[ht]
\centering
\subfigure[\ Mass vs radius. The observational limits imposed on maximum mass from the most massive pulsar PSR J0740+6620 \cite{Fonseca:2021wxt} and corresponding radius \cite{Miller:2021qha, Riley:2021pdl} are also indicated. The constraints on $M-R$ plane prescribed from GW170817 \cite{LIGOScientific:2018cki}, the NICER experiment for PSR J0030+0451 \cite{Riley:2019yda, Miller:2019cac}, HESS J1731-347 \cite{Doroshenko:2022}, PSR J0437-4715 \cite{Choudhury:2024xbk}, and PSR J1231-1411 \cite{Salmi:2024bss} are also compared.]{\includegraphics[width=0.48\linewidth,height=7cm]{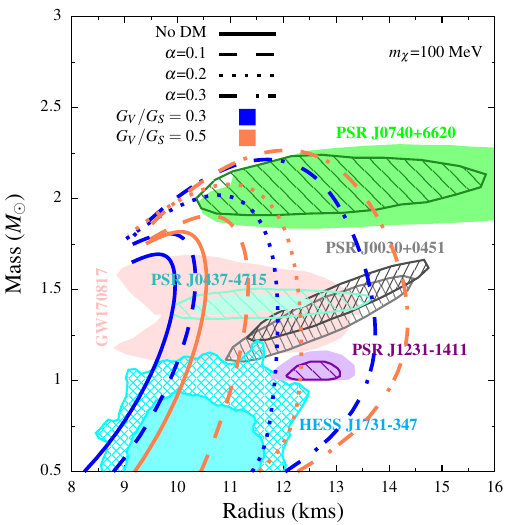} \label{fig:mr_Dnjl_mchi100M}}
\hfill 
\subfigure[\ Tidal deformability vs mass. The constraints on $\Lambda_{1.4}$ from GW170817 \cite{LIGOScientific:2018cki} is also shown.]{\includegraphics[width=0.48\linewidth,height=7cm]{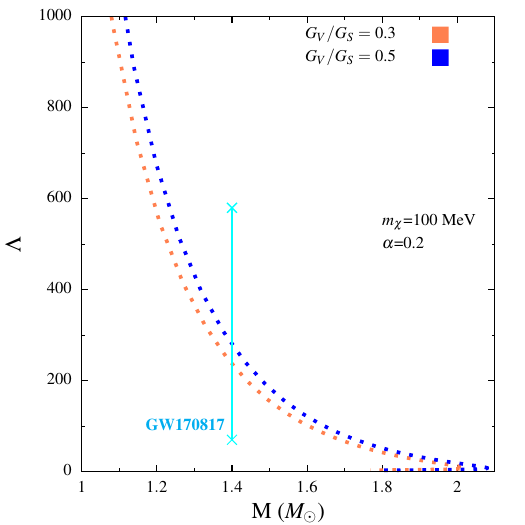}\label{fig:LamM_Dnjl_mchi100M}}
\caption{Variation of (a) mass with radius and (b) tidal deformability with mass of dark matter admixed strange quark stars for $G_V/G_S$=0.3, 0.5 and $m_{\chi}$=100 MeV and different values of $\alpha$.}
\label{fig:MR_Lam100M}
\end{figure*}
\begin{figure*}[ht]
\centering
\subfigure[\ Mass vs radius]{\includegraphics[width=0.48\linewidth,height=7cm]{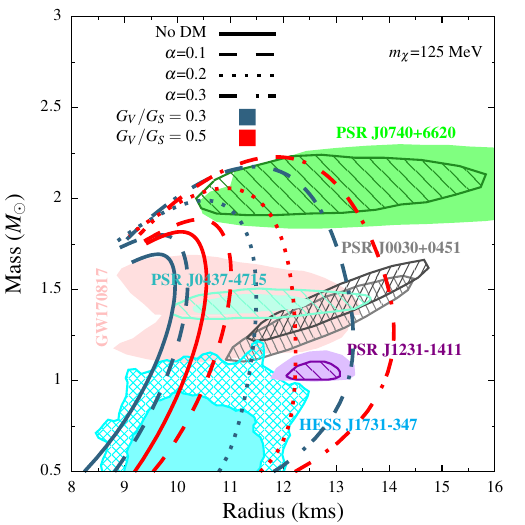} \label{fig:mr_Dnjl_mchi125M}}\quad
\hfill 
\subfigure[\ Tidal deformability vs mass]{\includegraphics[width=0.48\linewidth,height=7cm]{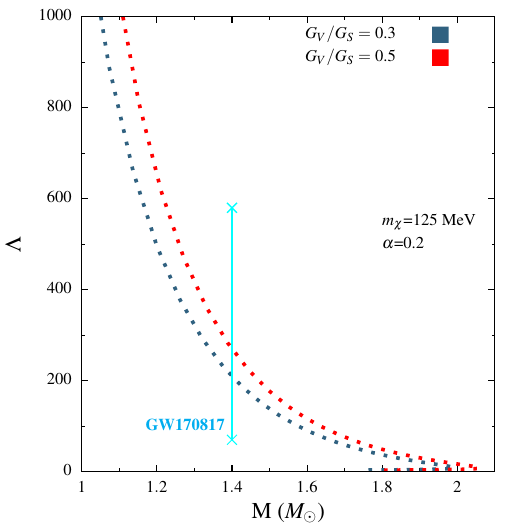}\label{fig:LamM_Dnjl_mchi125M}}
\caption{Variation of (a) mass with radius and (b) tidal deformability with mass of dark matter admixed strange quark stars for $G_V/G_S$=0.3, 0.5 and $m_{\chi}$=125 MeV and different values of $\alpha$.}
\label{fig:MR_Lam125M}
\end{figure*}
\begin{figure*}[ht]
\centering
\subfigure[\ Mass vs radius]{\includegraphics[width=0.48\linewidth,height=7cm]{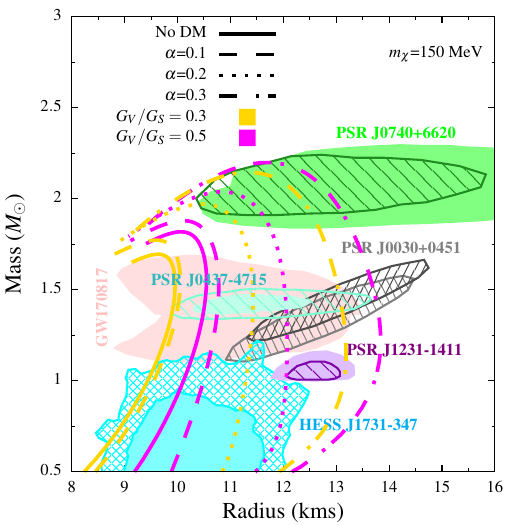} \label{fig:mr_Dnjl_mchi150M}}\quad
\hfill 
\subfigure[\ Tidal deformability vs mass]{\includegraphics[width=0.48\linewidth,height=7cm]{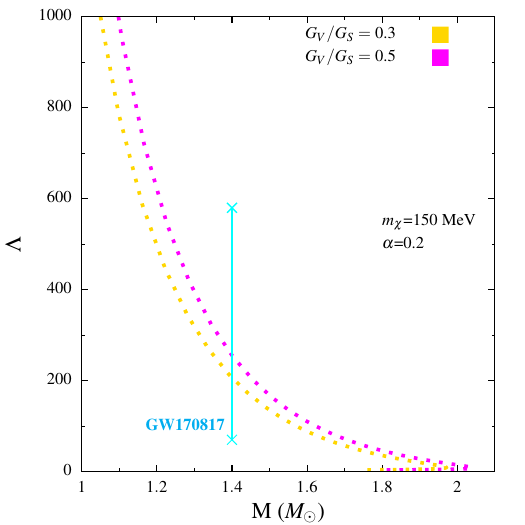}\label{fig:LamM_Dnjl_mchi150M}}
\caption{Variation of (a) mass with radius and (b) tidal deformability with mass of dark matter admixed strange quark stars for $G_V/G_S$=0.3, 0.5 and $m_{\chi}$=150 MeV and different values of $\alpha$.}
\label{fig:MR_Lam150M}
\end{figure*}
\subsection{Structural properties of dark matter admixed quark stars}
\label{subsec:Mass-radius-TD}
With the obtained EoS of the DMSQSs for the three benchmark values of $m_{\chi}$, we now study the structural properties of the DMSQSs like the mass and radius using the Eqs. (\ref{eq:tov}), (\ref{eq:tov2}) and (\ref{eq:tov3}) and also the tidal deformability using the Eq. (\ref{Eq:Lambda}). In Fig. \ref{fig:mr_Dnjl_mchi100M} we portray the variation of mass with respect to the radius of the DMSQSs for both $G_V/G_S$=0.3 and 0.5, with $m_{\chi}$=100 MeV. The mass of the DMSQSs increases with all the values of $\alpha$ compared to the no-DM case. The increasing values of $\alpha$ make the DMSQS substantially massive for a given value of $G_V/G_S$. For example, at $G_V/G_S$=0.3, the maximum mass of the DMSQS is $M_{max}$=1.81 $M_{\odot}$ for $\alpha$=0.1, $M_{max}$=2.02 $M_{\odot}$ for $\alpha$=0.2, and $M_{max}$=2.21 $M_{\odot}$ for $\alpha$=0.3. On the other hand, the radius ($R_{1.4}$) of 1.4 $M_{\odot}$ DMSQSs also increases largely with increasing values of $\alpha$. For example, at $G_V/G_S$=0.5, $R_{1.4}$=11.03 km for $\alpha$=0.1, $R_{1.4}$=12.31 km for $\alpha$=0.2, and $R_{1.4}$=14.35 km for $\alpha$=0.3. With $\alpha$=0.2 and 0.3, the maximum mass \cite{Fonseca:2021wxt} and the corresponding radius constraints \cite{Miller:2021qha, Riley:2021pdl} from PSR J0740+6620, are well satisfied. With $\alpha$=0.1, the mass-radius constraints from PSR J0740+6620, are not met for both $G_V/G_S$=0.3 and 0.5. For $G_V/G_S$=0.3 and 0.5, and $\alpha$=0.3, several other constraints on the $M-R$ plane such as those from GW170817 \cite{LIGOScientific:2018cki}, PSR J0437-4715 \cite{Choudhury:2024xbk}, and PSR J1231-1411 \cite{Salmi:2024bss}, are not fulfilled although the NICER data for PSR J0030+0451 \cite{Riley:2019yda, Miller:2019cac} is satisfied. Only for $\alpha$=0.2, all the aforementioned constraints along with that from HESS J1731-347 \cite{Doroshenko:2022}, are satisfied for both $G_V/G_S$=0.3 and 0.5. Since the values of $\alpha$ other than 0.2 are ruled out by the different $M-R$ constraints in Fig. \ref{fig:mr_Dnjl_mchi100M}, we display in Fig. \ref{fig:LamM_Dnjl_mchi100M} the variation of the tidal deformability with respect to the mass of the DMSQSs only for $\alpha$=0.2. For both $G_V/G_S$=0.3 and 0.5, the constraint on $\Lambda_{1.4}$ from GW170817 \cite{LIGOScientific:2018cki}, is harmoniously satisfied.

We repeat the same for other reference values of $m_{\chi}$=125 and 150 MeV in Figs. \ref{fig:MR_Lam125M} and \ref{fig:MR_Lam150M}, respectively. For both these values of $m_{\chi}$ and with $G_V/G_S$=0.3 and 0.5, we find that 0.2 is the most plausible value of $\alpha$ in light of the various recent constraints on the structural properties of compact stars. It can be observed from Figs. \ref{fig:mr_Dnjl_mchi100M}, \ref{fig:mr_Dnjl_mchi125M}, and \ref{fig:mr_Dnjl_mchi150M} that for any fixed values of $G_V/G_S$ and $\alpha$, the values of $M_{max}$ and $R_{1.4}$ of the DMSQSs show slight decrease with increasing values of $m_{\chi}$. For example, with $G_V/G_S$=0.5 and $\alpha$=0.2, $M_{max}$=2.08 $M_{\odot}$ for $m_{\chi}$=100 MeV, 2.06 $M_{\odot}$ for $m_{\chi}$=125 MeV, and 2.04 $M_{\odot}$ for $m_{\chi}$=150 MeV. Also with $G_V/G_S$=0.3 and $\alpha$=0.2, $R_{1.4}$=11.79 km for $m_{\chi}$=100 MeV, $R_{1.4}$=11.49 km for $m_{\chi}$=125 MeV, and $R_{1.4}$=11.43 km for $m_{\chi}$=150 MeV. The difference is not substantial because the three choices of $m_{\chi}$ are quite close, as seen from Fig. \ref{fig:mchi_LamSqD_LamVqD}. Thus for a given value of $G_V/G_S$, $\alpha$ affects $M_{max}$ and $R_{1.4}$ more than $m_{\chi}$.

\subsection{Oscillation properties of dark matter admixed quark stars}
\begin{figure*}[ht]
\centering
\subfigure[\ $m_{\chi}$=100 MeV]{\includegraphics[width=0.48\linewidth,height=7cm]{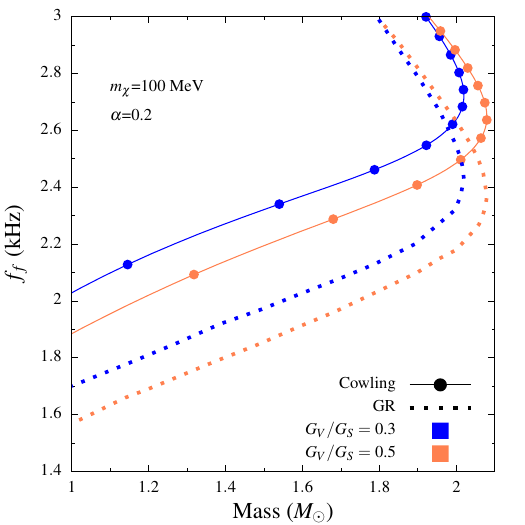}\label{fig:mf_Dnjl_mchi100M}}
\hfill 
\subfigure[\ $m_{\chi}$=125 MeV]{\includegraphics[width=0.48\linewidth,height=7cm]{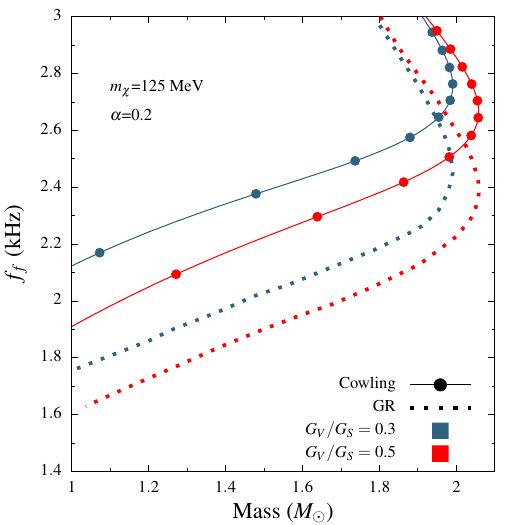}\label{fig:mf_Dnjl_mchi125M}}
\hfill 
\subfigure[\ $m_{\chi}$=150 MeV]{\includegraphics[width=0.48\linewidth,height=7cm]{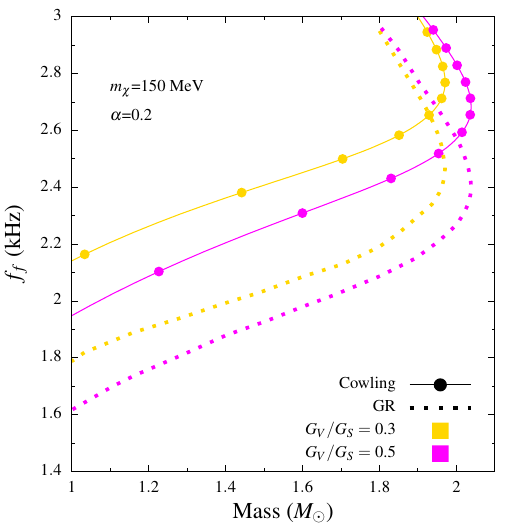}\label{fig:mf_Dnjl_mchi150M}}
\caption{Variation of $f$-mode frequency for dark matter admixed strange quark stars for $G_V/G_S$=0.3, 0.5 and (a) $m_{\chi}$=100 MeV and (b) $m_{\chi}$=125 MeV and (c) $m_{\chi}$=150 MeV.}
\label{fig:mf}
\end{figure*}
Once we obtain the values of the mass and radius of the DMSQSs, we next proceed to study the $f$-mode oscillation properties of the DMSQSs with $G_V/G_S$=0.3 and 0.5 and the three values of $m_{\chi}$, considering $\alpha$=0.2. For the calculation of the $f$-mode oscillation frequency ($f_f$) we adopt both Cowling approximation and GR approach as depicted in Sec. \ref{sec:Cowling} and \ref{sec:GR}, respectively. In Fig. \ref{fig:mf} we display the variation of $f$-mode frequency ($f_f$) with respect to mass of the DMSQSs. We show in Figs. \ref{fig:mf_Dnjl_mchi100M}, \ref{fig:mf_Dnjl_mchi125M}, and \ref{fig:mf_Dnjl_mchi150M} the results for $m_{\chi}$=100, 125, and 150 MeV, respectively, in both the Cowling approximation and GR calculations for $G_V/G_S$=0.3 and 0.5. The values of $f$-mode frequency in the GR calculation are always less than those in Cowling approximation. This is because the Cowling approximation often overestimates the value of $f_f$, especially at low mass, by almost up to 20$-$30\% \cite{Pradhan:2022vdf}. It is well known that compact stars with stiffer EoS oscillate with a lower $f$-mode frequency. Therefore, $f_f$ for $G_V/G_S$=0.3 is always greater than those for $G_V/G_S$=0.5. For example, at $m_{\chi}$=125 MeV and $\alpha$=0.2, $f_{f_{1.4}}$ (kHz) = 2.17 (1.84) with Cowling approximation (GR) when $G_V/G_S$=0.5 and $f_{f_{1.4}}$ (kHz) = 2.34 (1.97) when $G_V/G_S$=0.3. For the same reason with fixed values of $G_V/G_S$ and $\alpha$, $f_{f_{1.4}}$ decreases with increasing values of $m_{\chi}$. For example, at fixed $G_V/G_S$=0.3 and $\alpha$=0.2, $f_{f_{1.4}}$ (kHz) is 2.27 (1.92) when $m_{\chi}$=100 MeV, 2.34 (1.97) when $m_{\chi}$=125 MeV, and 2.37 (1.99) when $m_{\chi}$=150 MeV obtained with Cowling approximation (GR). Therefore, it is also easy to understand that for fixed values of $G_V/G_S$ and $m_{\chi}$, $f_{f_{1.4}}$ will decrease with increasing values of $\alpha$.
\begin{figure*}[ht]
\centering
\subfigure[\ $f_{f_{1.4}}$ vs $R_{1.4}$]{\includegraphics[width=0.48\linewidth,height=7cm]{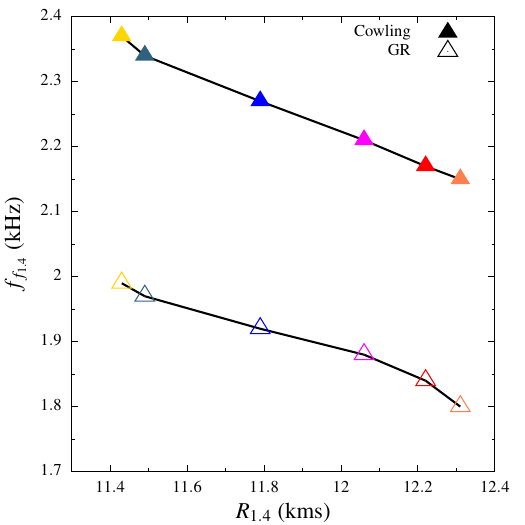}\label{fig:f1p4R1p4}}
\hfill 
\subfigure[\ $f_{f_{1.4}}$ vs $\Lambda_{1.4}$]{\includegraphics[width=0.48\linewidth,height=7cm]{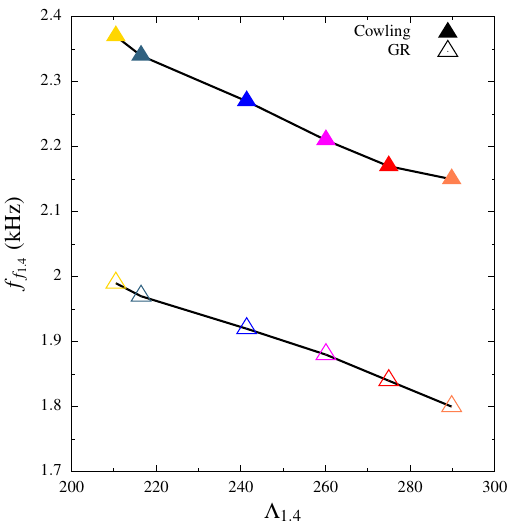}\label{fig:f1p4Lam1p4}}
\caption{Variation of $f$-mode frequency with (a) radius and (b) tidal deformability of 1.4 $M_{odot}$ dark matter admixed strange quark stars. The color of the points corresponds to the color of the lines in Figs. \ref{fig:mf_Dnjl_mchi100M}, \ref{fig:mf_Dnjl_mchi125M}, and \ref{fig:mf_Dnjl_mchi150M}.}
\label{fig:f1p4_R1p4_Lam1p4}
\end{figure*}

\subsubsection*{Universal relations}

Several works suggest that, irrespective of the composition and EoS of the compact stars, the $f$-mode frequency of 1.4 $M_{\odot}$ star is found to be well correlated with both $R_{1.4}$ and $\Lambda_{1.4}$. Therefore, we test the same in the case of DMSQSs in Fig. \ref{fig:f1p4_R1p4_Lam1p4}. The universal relations $f_{f_{1.4}}-R_{1.4}$ and $f_{f_{1.4}}-\Lambda_{1.4}$ are studied in Figs. \ref{fig:f1p4R1p4} and \ref{fig:f1p4Lam1p4}, respectively. In either case, negative correlation is obtained and this result is found to be consistent with other works in the literature. In both Cowling approximation and GR treatment, the decrement of $f_{f_{1.4}}$ with both $R_{1.4}$ and $\Lambda_{1.4}$ is found to be almost linear. The $f_{f_{1.4}}-R_{1.4}$ and $f_{f_{1.4}}-\Lambda_{1.4}$ variations obtained with the Cowling approximation and GR calculation differ substantially because the values of $f_{f_{1.4}}$ are quite less in the case of GR treatment compared to those obtained with Cowling approximation as seen from Fig. \ref{fig:mf}.

\begin{figure*}[ht]
\centering
{\includegraphics[width=0.6\linewidth,height=8cm]{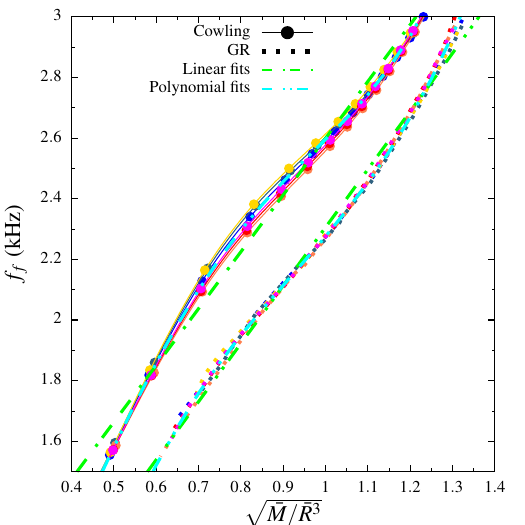}}
\caption{Variation of $f$-mode frequency with average density of dark matter admixed strange quark stars.}
\label{fig:f_rhobar}
\end{figure*}

\begin{figure*}[ht]
\centering
\subfigure[\ mass-scaled $f$-mode frequency vs compactness]
{\includegraphics[width=0.48\linewidth,height=7cm]{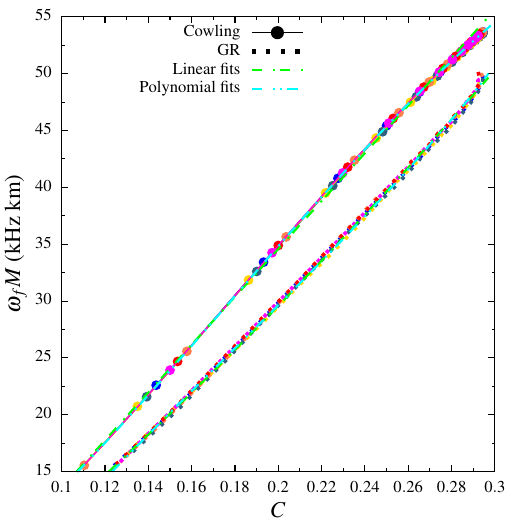}\label{fig:omgM_C}}
\hfill
\subfigure[\ $f$-mode frequency vs tidal deformability]{\includegraphics[width=0.48\linewidth,height=7cm]{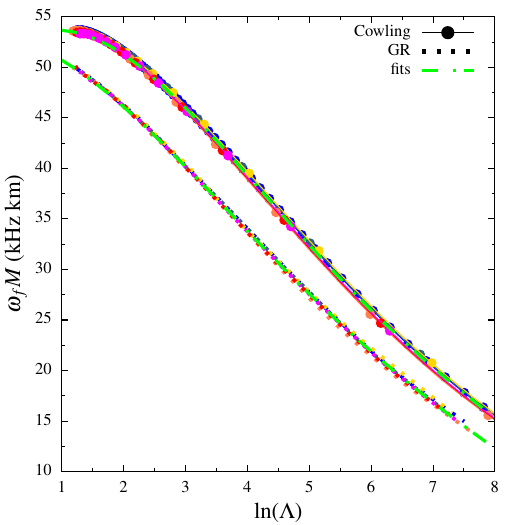}\label{fig:omgM_Lam}}
\label{fig:f_rhobar_omgM_Lam}
\caption{Variation of mass-scaled $f$-mode frequency with (a) compactness and (b) tidal deformability of dark matter admixed strange quark stars.}
\end{figure*}

We next focus on the other universal relations related to the $f$-mode oscillation frequency in terms of average density, compactness, and tidal deformability, calculated for the DMSQSs with our DNJL model. In Fig. \ref{fig:f_rhobar} we plot $f_f$ with respect to the average density $\bar{\rho}=\sqrt{\bar{M}/\bar{R}^3}$ of DMSQSs. It can be seen that the $f$-mode frequency with respect to $\bar{\rho}$ shows universality in both the Cowling approximation and the GR calculation. Similarly, Figs. \ref{fig:omgM_Lam} and \ref{fig:omgM_C} suggest that the mass scaled angular frequency $\omega_f M$ (corresponding to the $f$-mode oscillation) holds universality with quantities like compactness $C=M/R$ and the tidal deformability of DMSQSs, respectively in both the Cowling approximation and the GR calculation. For all three cases in Figs. \ref{fig:f_rhobar}, \ref{fig:omgM_C} and \ref{fig:omgM_Lam} the universal relations are obtained irrespective of the values of $m_{\chi}$, $G_V/G_S$, and $\alpha$. These relations deviate noticeably from the Cowling approximation to the GR calculation. For the $f_f-\bar{\rho}$ variation in Fig. \ref{fig:f_rhobar}, the polynomial fitted relation is
\begin{eqnarray}
f_f {\rm{(kHz)}} = \sum_i a_i \bar{\rho}^i
\label{eq:fdens_polynomial}
\end{eqnarray}
where, $i$=0,1,2,3 and for the polynomial fit $a_0 = -1.332~(-2.024)$, $a_1 = 9.050~(10.397)$, $a_2 = 2.693~(-9.551)$, and $a_3 = 2.693~(3.454)$ in the Cowling approximation (GR). 

For the $\omega_f M-C$ plot in Fig. \ref{fig:omgM_C}, the polynomial fit is given as
\begin{eqnarray}
\omega_f M {\rm{(kHzkm)}} = \sum_i a_i C^i
\label{eq:omgC_polynomial}
\end{eqnarray}
where, $i$=0,1,2,3 and $a_0=-1.568~(-14.371)$, $a_1=93.735~(295.261)$, $a_2=701.438~(-603.166)$, and $a_3=-1301.284~(1136.716)$ in the Cowling approximation (GR). As seen in Figs. \ref{fig:f_rhobar} and \ref{fig:omgM_C}, the polynomial fits are more accurate than the linear fits in both cases. However, some works have presented linear fits for these two universal relations considering different compositions of compact stars. Therefore, we also compare the linear fits in Figs. \ref{fig:f_rhobar} and \ref{fig:omgM_C}. Consequently, we compare the coefficients of our linear fits in Eqs. (\ref{eq:fdens_polynomial}) and (\ref{eq:omgC_polynomial}) up to $i$=1 in both Cowling approximation and GR technique with other works done for various compositions of compact stars in Tabs. \ref{tab:fdens_polynomial} and \ref{tab:omgC_polynomial}, respectively.

\begin{table}[!ht]
\caption{The coefficients of the linear fits viz. $a_0$ and $a_1$ (in kHz) corresponding to Fig. \ref{fig:f_rhobar} and Eq. (\ref{eq:fdens_polynomial}) calculated with Cowling approximation and GR.}
\label{tab:fdens_polynomial}
{{
\setlength{\tabcolsep}{20pt}
\begin{center}
\begin{tabular}{c| c c c c}
\hline
 & \multicolumn{2}{c|}{Cowling} & \multicolumn{2}{c}{GR}   \\
\hline

Refs. & \multicolumn{1}{c|}{$a_0$} & \multicolumn{1}{c|}{$a_1$} & \multicolumn{1}{c|}{$a_0$} & \multicolumn{1}{c}{$a_1$} \\

\cline{1-5}
 This work (SGP) & \multicolumn{1}{c|}{0.727} & \multicolumn{1}{c|}{1.872} & \multicolumn{1}{c|}{0.391} & \multicolumn{1}{c}{1.914} \\

DGKK \cite{Doneva:2013zqa} & \multicolumn{1}{c|}{1.562} & \multicolumn{1}{c|}{1.151}  & \multicolumn{1}{c|}{-} & \multicolumn{1}{c}{-}  \\
 
JC \cite{Jaiswal:2020wzu} & \multicolumn{1}{c|}{0.6499, 0.6397} & \multicolumn{1}{c|}{1.6248, 1.6243} & \multicolumn{1}{c|}{-} & \multicolumn{1}{c}{-} \\ 
 
DKBP \cite{Das:2021dru} & \multicolumn{1}{c|}{1.185, 1.256} & \multicolumn{1}{c|}{1.246, 1.311} & \multicolumn{1}{c|}{-} & \multicolumn{1}{c}{-} \\ 

KMMP \cite{Kumar:2023rut} & \multicolumn{1}{c|}{1.520, 1.348} & \multicolumn{1}{c|}{0.833, 1.001} & \multicolumn{1}{c|}{-} & \multicolumn{1}{c}{-} \\

JNSGS \cite{Jyothilakshmi:2024zqn} & \multicolumn{1}{c|}{1.367} & \multicolumn{1}{c|}{0.87} & \multicolumn{1}{c|}{-} & \multicolumn{1}{c}{-} \\

HRWM \cite{Hong:2023udv} & \multicolumn{1}{c|}{1.0581} & \multicolumn{1}{c|}{1.4507} & \multicolumn{1}{c|}{-} & \multicolumn{1}{c}{-} \\

PC \cite{Pradhan:2020amo} & \multicolumn{1}{c|}{1.075} & \multicolumn{1}{c|}{1.412} & \multicolumn{1}{c|}{-} & \multicolumn{1}{c}{-} \\

DPPBP \cite{Dey:2024vsw} & \multicolumn{1}{c|}{0.628} & \multicolumn{1}{c|}{2.048} & \multicolumn{1}{c|}{-} & \multicolumn{1}{c}{-} \\ 

RMTL \cite{Rather:2024mtd} & \multicolumn{1}{c|}{1.32, 1.29} & \multicolumn{1}{c|}{1.18, 1.22} & \multicolumn{1}{c|}{0.44, 0.39} & \multicolumn{1}{c}{1.72, 1.79} \\ 

BFG \cite{Benhar:2004xg} & \multicolumn{1}{c|}{-} & \multicolumn{1}{c|}{-} & \multicolumn{1}{c|}{0.79} & \multicolumn{1}{c}{1.235} \\

AK \cite{Andersson:1996pn} & \multicolumn{1}{c|}{-} & \multicolumn{1}{c|}{-} & \multicolumn{1}{c|}{0.78} & \multicolumn{1}{c}{1.635} \\

PCLJ \cite{Pradhan:2022vdf} & \multicolumn{1}{c|}{-} & \multicolumn{1}{c|}{-} & \multicolumn{1}{c|}{0.535} & \multicolumn{1}{c}{1.354} \\

SPCCSS-B \cite{Shirke:2024ymc} & \multicolumn{1}{c|}{-} & \multicolumn{1}{c|}{-} & \multicolumn{1}{c|}{0.535} & \multicolumn{1}{c}{1.354} \\

CKK \cite{Celato:2025klx} & \multicolumn{1}{c|}{-} & \multicolumn{1}{c|}{-} & \multicolumn{1}{c|}{-0.01461, 0.06661} & \multicolumn{1}{c}{1.576, 1.458} \\ 

\hline

\end{tabular}
\end{center}
}}
\end{table} 

\begin{table}[!ht]
\caption{The coefficients of the linear fits viz. $a_0$ and $a_1$ (in kHzkm) corresponding to Fig. \ref{fig:omgM_Lam} and Eq. (\ref{eq:omgC_polynomial}) calculated with Cowling approximation. In case of GR, the value of $a_2$ (in kHzkm) is also reported and compared in addition to $a_0$ and $a_1$.}
\label{tab:omgC_polynomial}
{{
\setlength{\tabcolsep}{15pt}
\begin{center}
\begin{tabular}{c| c c c c c}
\hline
& \multicolumn{2}{c|}{Cowling} & \multicolumn{3}{c}{GR}   \\
\hline

Refs. & \multicolumn{1}{c|}{$a_0$} & \multicolumn{1}{c|}{$a_1$} & \multicolumn{1}{c|}{$a_0$} & \multicolumn{1}{c|}{$a_1$} & \multicolumn{1}{c}{$a_2$} \\

\cline{1-6}
This work (SGP) & \multicolumn{1}{c|}{-7.479} & \multicolumn{1}{c|}{201.310} & \multicolumn{1}{c|}{-4.834} & \multicolumn{1}{c|}{148.654} & \multicolumn{1}{c}{117.163} \\
 
DKBP \cite{Das:2021dru} & \multicolumn{1}{c|}{-4.538, -2.984} & \multicolumn{1}{c|}{190.447, 190.475} & \multicolumn{1}{c|}{-} & \multicolumn{1}{c|}{-} & \multicolumn{1}{c}{-} \\ 
 
PC \cite{Pradhan:2020amo} & \multicolumn{1}{c|}{-3.836} & \multicolumn{1}{c|}{197.295} & \multicolumn{1}{c|}{-} & \multicolumn{1}{c|}{-} & \multicolumn{1}{c}{-} \\ 

GSH \cite{Guha:2024gfe} & \multicolumn{1}{c|}{-5.65} & \multicolumn{1}{c|}{200.41} & \multicolumn{1}{c|}{-} & \multicolumn{1}{c|}{-} & \multicolumn{1}{c}{-} \\ 

JNSGS \cite{Jyothilakshmi:2024zqn} & \multicolumn{1}{c|}{-5.285} & \multicolumn{1}{c|}{190.390} & \multicolumn{1}{c|}{-} & \multicolumn{1}{c|}{-} & \multicolumn{1}{c}{-} \\

KMMP \cite{Kumar:2023rut} & \multicolumn{1}{c|}{-2.552, 3.093} & \multicolumn{1}{c|}{194.414, 196.67} & \multicolumn{1}{c|}{-} & \multicolumn{1}{c|}{-} & \multicolumn{1}{c}{-} \\

DPPBP \cite{Dey:2024vsw} & \multicolumn{1}{c|}{-4.665} & \multicolumn{1}{c|}{199.95} & \multicolumn{1}{c|}{-} & \multicolumn{1}{c|}{-} & \multicolumn{1}{c}{-} \\

RMTL \cite{Rather:2024mtd} & \multicolumn{1}{c|}{-3.66, -3.88} & \multicolumn{1}{c|}{199.40, 200.0} & \multicolumn{1}{c|}{6.63, -6.92} & \multicolumn{1}{c|}{179.61, 180.65} & \multicolumn{1}{c}{-} \\

TL \cite{Tsui:2004qd} & \multicolumn{1}{c|}{-} & \multicolumn{1}{c|}{-} & \multicolumn{1}{c|}{-6.01} & \multicolumn{1}{c|}{168.168} & \multicolumn{1}{c}{23.724} \\

PCA-C \cite{Pradhan:2023zmg} & \multicolumn{1}{c|}{-} & \multicolumn{1}{c|}{-}  & \multicolumn{1}{c|}{-7.207} & \multicolumn{1}{c|}{176.876} & \multicolumn{1}{c}{26.727} \\

PCLJ \cite{Pradhan:2022vdf} & \multicolumn{1}{c|}{-} & \multicolumn{1}{c|}{-} & \multicolumn{1}{c|}{-7.808} & \multicolumn{1}{c|}{179.880} & \multicolumn{1}{c}{23.724} \\

WLCZ \cite{Wen:2019ouw} & \multicolumn{1}{c|}{-} & \multicolumn{1}{c|}{-} & \multicolumn{1}{c|}{-6.01} & \multicolumn{1}{c|}{168.168} & \multicolumn{1}{c}{45.045} \\

\hline

\end{tabular}
\end{center}
}}
\end{table} 

For the $\omega_f M-{\rm{ln}}~(\Lambda)$ relation shown in Fig. \ref{fig:omgM_Lam}, similar to other works, we have considered only polynomial fits, which are given as
\begin{eqnarray}
\omega_f M {\rm{(kHzkm)}} = \sum_j \alpha_j ({\rm{ln}}~\Lambda)^j,
\label{eq:omgLam_polynomial}
\end{eqnarray}
We consider the coefficients of $({\rm{ln}}~\Lambda)^j$ up to $j$=4 in Eq. (\ref{eq:omgLam_polynomial}) and compare them with those obtained in other works in Cowling approximation and GR separately in Tab. \ref{tab:omgLam_polynomial}.
\begin{table}[!ht]
\caption{The coefficients of the fits viz. $a_j$s (in kHzkm) corresponding to Fig. \ref{fig:omgM_Lam} and Eq. (\ref{eq:omgLam_polynomial}) calculated with Cowling approximation and GR.}
\label{tab:omgLam_polynomial}
{{
\setlength{\tabcolsep}{3pt}
\begin{center}
\begin{tabular}{c| c c c c c c c c c c c c c c}
\hline
& \multicolumn{7}{c|}{Cowling} & \multicolumn{7}{c}{GR}   \\
\hline

Refs. & \multicolumn{1}{c|}{$\alpha_0$} & \multicolumn{1}{c|}{$\alpha_1$} & \multicolumn{1}{c|}{$\alpha_2$} & \multicolumn{1}{c|}{$\alpha_3$} & \multicolumn{1}{c|}{$\alpha_4$} & \multicolumn{1}{c|}{$\alpha_5$} & \multicolumn{1}{c|}{$\alpha_6$} & \multicolumn{1}{c|}{$\alpha_0$} & \multicolumn{1}{c|}{$\alpha_1$} & \multicolumn{1}{c|}{$\alpha_2$} & \multicolumn{1}{c|}{$\alpha_3$} & \multicolumn{1}{c|}{$\alpha_4$} & \multicolumn{1}{c|}{$\alpha_5$} & \multicolumn{1}{c}{$\alpha_6$} \\

\cline{1-15}
This work (SGP) & \multicolumn{1}{c|}{51.296} & \multicolumn{1}{c|}{5.485} & \multicolumn{1}{c|}{-3.520} & \multicolumn{1}{c|}{0.416} & \multicolumn{1}{c|}{-0.0165} & \multicolumn{1}{c|}{-} & \multicolumn{1}{c|}{-} & \multicolumn{1}{c|}{52.854} & \multicolumn{1}{c|}{-0.494} & \multicolumn{1}{c|}{-1.899} & \multicolumn{1}{c|}{0.250} & \multicolumn{1}{c|}{-0.0105} & \multicolumn{1}{c|}{-} & \multicolumn{1}{c}{-} \\
 
JNSGS \cite{Jyothilakshmi:2024zqn} & \multicolumn{1}{c|}{57.057} & \multicolumn{1}{c|}{1.201} & \multicolumn{1}{c|}{-1.847} & \multicolumn{1}{c|}{0.197} & \multicolumn{1}{c|}{-0.008} & \multicolumn{1}{c|}{0.0001} & \multicolumn{1}{c|}{-} & \multicolumn{1}{c|}{-} & \multicolumn{1}{c|}{-} & \multicolumn{1}{c|}{-} & \multicolumn{1}{c|}{-} & \multicolumn{1}{c|}{-} & \multicolumn{1}{c|}{-} & \multicolumn{1}{c}{-} \\ 
 
KGTTNS \cite{Kumar:2023ojk} & \multicolumn{1}{c|}{59.459} & \multicolumn{1}{c|}{-0.793} & \multicolumn{1}{c|}{-1.471} & \multicolumn{1}{c|}{0.166} & \multicolumn{1}{c|}{-0.007} & \multicolumn{1}{c|}{0.00008} & \multicolumn{1}{c|}{-} & \multicolumn{1}{c|}{55.255} & \multicolumn{1}{c|}{-1.325} & \multicolumn{1}{c|}{-1.621} & \multicolumn{1}{c|}{0.235} & \multicolumn{1}{c|}{-0.013} & \multicolumn{1}{c|}{0.0003} & \multicolumn{1}{c}{-} \\ 

CSLL \cite{Chan:2014kua} & \multicolumn{1}{c|}{-} & \multicolumn{1}{c|}{-} & \multicolumn{1}{c|}{-} & \multicolumn{1}{c|}{-} & \multicolumn{1}{c|}{-} & \multicolumn{1}{c|}{-} & \multicolumn{1}{c|}{-} & \multicolumn{1}{c|}{54.655} & \multicolumn{1}{c|}{-2.053} & \multicolumn{1}{c|}{0.157} & \multicolumn{1}{c|}{-0.005} & \multicolumn{1}{c|}{-} & \multicolumn{1}{c|}{-} & \multicolumn{1}{c}{-} \\

PCLJ \cite{Pradhan:2022vdf} & \multicolumn{1}{c|}{-} & \multicolumn{1}{c|}{-} & \multicolumn{1}{c|}{-} & \multicolumn{1}{c|}{-} & \multicolumn{1}{c|}{-} & \multicolumn{1}{c|}{-} & \multicolumn{1}{c|}{-} & \multicolumn{1}{c|}{54.474} & \multicolumn{1}{c|}{-1.749} & \multicolumn{1}{c|}{-1.419} & \multicolumn{1}{c|}{0.190} & \multicolumn{1}{c|}{-0.086} & \multicolumn{1}{c|}{0.00009} & \multicolumn{1}{c}{-} \\

PVC \cite{Pradhan:2022rxs} & \multicolumn{1}{c|}{-} & \multicolumn{1}{c|}{-} & \multicolumn{1}{c|}{-} & \multicolumn{1}{c|}{-} & \multicolumn{1}{c|}{-} & \multicolumn{1}{c|}{-} & \multicolumn{1}{c|}{-} & \multicolumn{1}{c|}{54.655} & \multicolumn{1}{c|}{-2.001} & \multicolumn{1}{c|}{-1.265} & \multicolumn{1}{c|}{0.142} & \multicolumn{1}{c|}{-0.0003} & \multicolumn{1}{c|}{-0.0006} & \multicolumn{1}{c}{0.00003} \\

SNG \cite{Shashank:2021oxe} & \multicolumn{1}{c|}{-} & \multicolumn{1}{c|}{-} & \multicolumn{1}{c|}{-} & \multicolumn{1}{c|}{-} & \multicolumn{1}{c|}{-} & \multicolumn{1}{c|}{-} & \multicolumn{1}{c|}{-} & \multicolumn{1}{c|}{43.303} & \multicolumn{1}{c|}{9.024} & \multicolumn{1}{c|}{-4.826} & \multicolumn{1}{c|}{0.628} & \multicolumn{1}{c|}{-0.028} & \multicolumn{1}{c|}{-} & \multicolumn{1}{c}{-} \\
\hline
\end{tabular}
\end{center}
}}
\end{table} 

Overall, Tabs. \ref{tab:fdens_polynomial}, \ref{tab:omgC_polynomial}, and \ref{tab:omgLam_polynomial} show that our fits for the DMSQSs are quite comparable with other works in literature, obtained with different compositions of the compact stars. These universal relations obtained with our DMSQSs suggest that we have obtained reasonable DMSQS configurations with respect to not only the structural but also the oscillation properties of the compact stars.

\subsubsection*{Observational prospect}
\begin{figure*}[ht]
\centering
\subfigure[\ $f$-mode frequency vs $f$-mode damping time]{\includegraphics[width=0.48\linewidth,height=7cm]{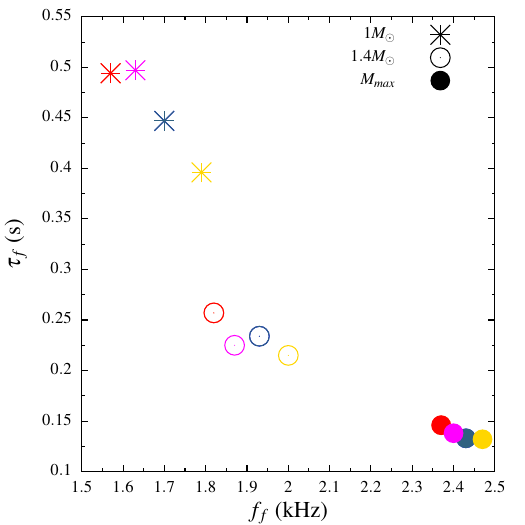}\label{fig:ftau_Dnjl_damping}}
\hfill 
\subfigure[\ amplitude of the gravitational wave strain vs $f$-mode frequency]
{\includegraphics[width=0.49\linewidth,height=7cm]{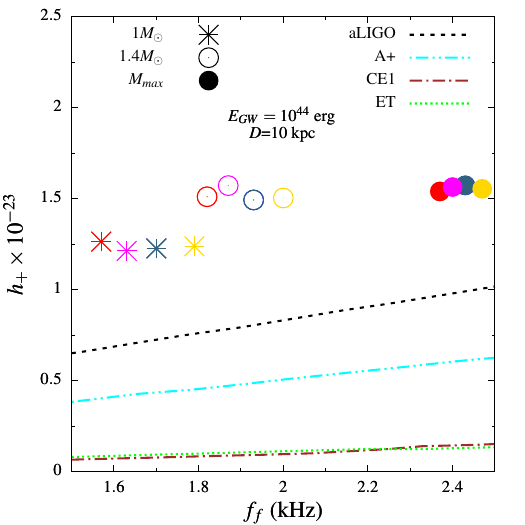}\label{fig:fhplus_Dnjl}}
\caption{Variation of $f$-mode frequency with (a) $f$-mode damping time, (b) amplitude of the gravitational wave strain of dark matter admixed strange quark stars of fixed stellar mass $1 M_{\odot}$, $1.4 M_{\odot}$, and maximum mass. The sensitivities of Advanced LIGO (aLIGO) \cite{KAGRA:2013rdx}, A+ \cite{A+}, Cosmic Explorer (CE1) \cite{CE1}, and Einstein Telescope (ET) \cite{Hild:2010id, Punturo:2010zz} are also compared in (b).}
\label{fig:f_f_tau_strain}
\end{figure*}

Considering the glitch energy that is emitted by GW radiation via $f$-mode oscillation as $E_{GW}=10^{44}$ ergs, we calculate the GW strain amplitude $h_+$ for a pulsar at distance $D=10$ kpc (e.g. in the Messier 3 (M3 or NGC 5272) globular cluster) with the DMSQS EoS for fixed mass 1$M_{\odot}$, 1.4$M_{\odot}$ and at maximum mass using Eq. (\ref{eq:strain}). For this purpose, we need to know the damping time ($\tau_f$) of the $f$-mode oscillation frequency. Thus, the calculation of $h_+$ is only possible within GR approach. In Fig. \ref{fig:ftau_Dnjl_damping} we present the variation of the damping time with respect to frequency of the $f$-mode oscillation of the DMSQSs of fixed mass mentioned before. It can be seen that the damping time is more for stars that oscillate with low frequency, indicating a negative correlation between $\tau_f$ and $f_f$. There is a very distinct separation of the values $\tau_f$ (with respect to $f_f$) for the different masses of DMSQSs. The most massive stars with maximum mass (indicated with solid points) oscillate with maximum frequency and least damping time, followed by the 1.4$M_{\odot}$ DMSQSs (shown with open points) oscillate with intermediate frequency and damping time. The 1$M_{\odot}$ DMSQSs (shown with asterisks) oscillate with the least frequency and the maximum damping time. This feature is also reflected in the corresponding evaluation of the amplitude of the GW strain in Fig. \ref{fig:fhplus_Dnjl} with respect to $f_f$. The calculated values of $h_+$ for the three masses of the DMSQSs fall within the detection capabilities of the various upcoming GW detectors like aLIGO, A+, CE1, and ET, as understood from their projected sensitivities. However, $h_+$ depends on $E_{GW}$ and $D$. From Eq. (\ref{eq:strain}), we find that for $E_{GW}\sim10^{42}-10^{44}$ ergs and $D\sim1-10$ kpc (e.g. Jewel Box (NGC 4755) and 47 Tucanae (NGC 104)), detection is possible by all the four upcoming GW detectors.

\subsection{Bayesian analysis of the parameters $G_V/G_S$, $\alpha$, and $\rho_{sc}$ in the light of existing observational constraints}

\begin{figure*}[ht]
\centering
\begin{picture}(10,9)
\put(10,5){\LARGE{$m_\chi = 100~\rm{MeV}$}}
\end{picture}
{\includegraphics[width=1.0\linewidth]{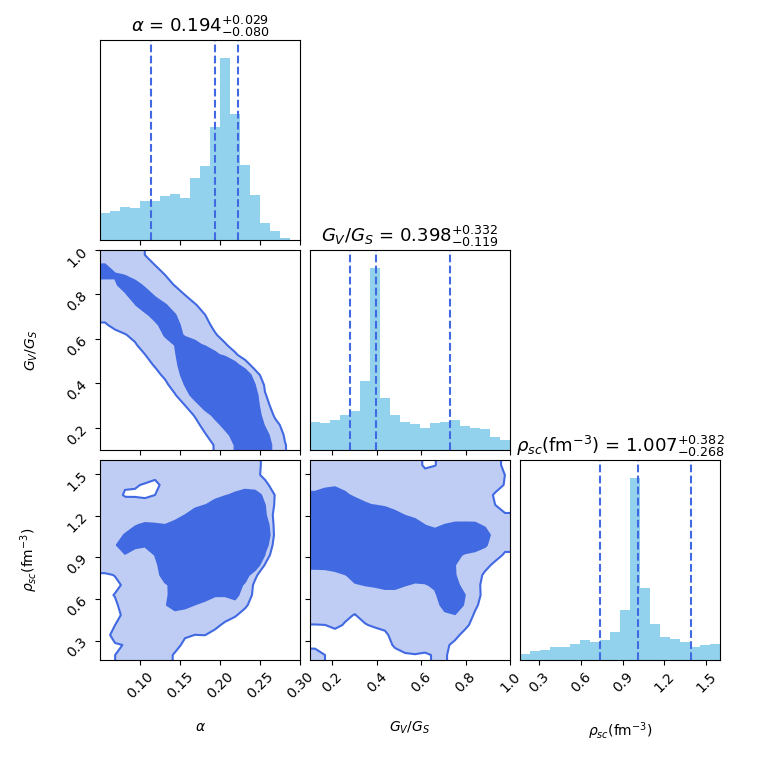}}
\caption{Bayesian analysis of the free parameters $\alpha$, $G_V/G_S$ and $\rho_{sc}$ for $m_\chi =100~\rm{MeV}$.}
\label{fig:corner_plot_100}
\end{figure*}
\begin{figure*}[ht]
\centering
\begin{picture}(10,9)
\put(10,5){\LARGE{$m_\chi = 150~\rm{MeV}$}}
\end{picture}
{\includegraphics[width=1.0\linewidth]{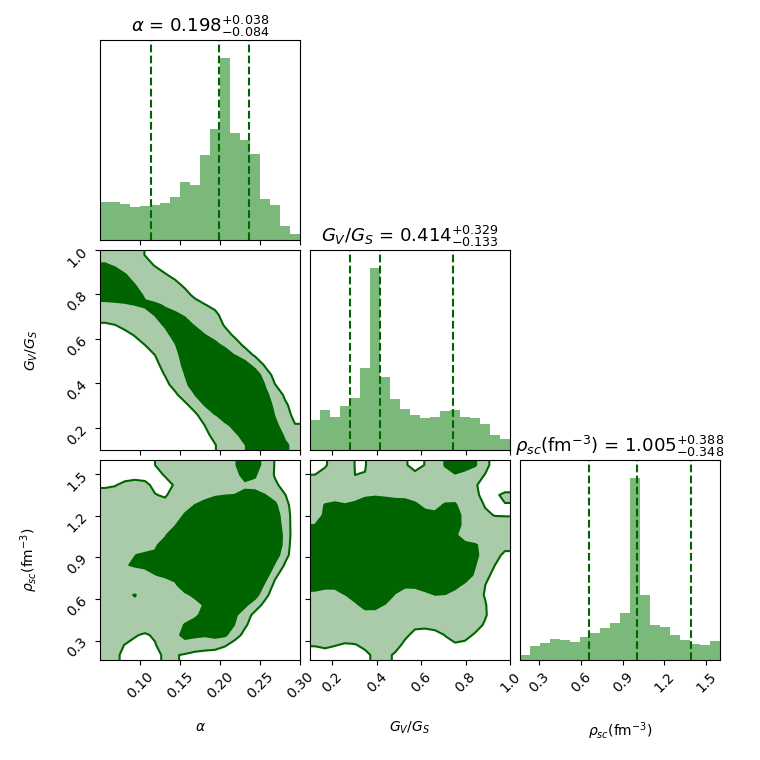}}
\caption{Bayesian analysis of the free parameters $\alpha$, $G_V/G_S$ and $\rho_{sc}$ for $m_\chi =150~\rm{MeV}$.}
\label{fig:corner_plot_150}
\end{figure*}
The formulation of the Bayesian analysis, making use of the recent observational constraints, has been discussed in Sec. \ref{Sec:Bayesian analysis}. The results obtained in Sec. \ref{subsec:DMQM} and \ref{subsec:Mass-radius-TD} guide us to set proper prior distributions for the three free parameters of our DNJL model, viz. $G_V/G_S$, $\alpha$, and $\rho_{sc}$. We set the uniform prior distributions of these parameters within the following range.

\begin{itemize}
\centering
\item Range for $\alpha$ = $0.1-0.3$
\item Range for $G_V/G_S$ = $0.3-0.5$
\item Range for $\rho_{sc}$ = $\rho_0-10\rho_0$ = $0.16$-$1.6~\rm{fm^{-3}}$ (where, $\rho_0$ is the nuclear saturation density)
\end{itemize}    

After performing the Bayesian analysis, we obtain the posterior distribution of the parameters. We show the distribution along with the $68\%$ and $95\%$ credible intervals in Figs. \ref{fig:corner_plot_100} and \ref{fig:corner_plot_150} for $m_{\chi}=$ 100 MeV and 150 MeV, respectively. The most credible values and Maximum A Posteriori (MAP) estimates for the parameters are listed below. For $m_\chi = 100$ MeV, we obtain the following.

\begin{itemize}
\centering
\item Most credible $\alpha$ = $0.194^{+0.029}_{-0.080}$ ; MAP of $\alpha$ = $0.201$
\item Most credible $G_V/G_S$ = $0.398^{+0.332}_{-0.119}$ ; MAP of $G_V/G_S$ = $0.351$
\item Most credible $\rho_{sc}\rm{(fm^{-3})}$ = $1.007^{+0.382}_{-0.268}$ ; MAP of $\rho_{sc}\rm{(fm^{-3})}$ = $0.998$.
\end{itemize}

and for $m_\chi = 150$ MeV the estimates are as follows.

\begin{itemize}
\centering
\item Most credible $\alpha$ = $0.198^{+0.038}_{-0.084}$ ; MAP of $\alpha$ = $0.212$
\item Most credible $G_V/G_S$ = $0.414^{+0.329}_{-0.133}$ ; MAP of $G_V/G_S$ = $0.346$
\item Most credible $\rho_{sc}\rm{(fm^{-3})}$ = $1.005^{+0.388}_{-0.348}$ ; MAP of $\rho_{sc}\rm{(fm^{-3})}$ = $0.999$.
\end{itemize}

The above results show that the constraints on the three parameters are quite stringent and conservative with respect to $m_{\chi}$ in case of contact type interaction between SQM and DM of mass in sub-GeV range.

\section{Summary and Conclusion}
\label{Conclusion}

This work is dedicated to the study of the structural and oscillation properties of DMSQSs. For the purpose, we consider the NJL model with the HK parameterization for the SQM. The dark fermionic sector is self-interacting via contact-type four-Fermi interaction. The DM-SQM interaction is also invoked by considering contact-type four-Fermi interaction. The DM density is related to the baryon density of SQM via a parameterized exponential form with two free parameters viz., $\alpha$ and $\rho_{sc}$. In case of four-Fermi type of interaction, in absence of mediators, the DM mass is chosen systematically in par with the results of direct detection experiments like LZ, XENON, DarkSide, CRESST, and the collider experiment LHC, and also the bound from relic abundance. These experiments have set strong constraints on the choice of $m_{\chi}$ and the momentum cut-offs in the case of contact-type of interaction between DM and SQM. They indicate that such type of interaction is almost impossible with heavy DM ($m_{\chi}\geq$ 1 GeV). In the present work, we show that, apart from these experiments, the existing astrophysical constraints on the structural properties of compact stars also deplore the presence of heavy DM interacting with SQM in DMSQSs. Therefore, in the recent era, the hunt for sub-GeV DM is gaining importance. In accordance with such circumstances, we emphasize that light DM (of mass in MeV scale) can possibly exist in DMSQSs and interact with SQM via contact type of interaction. This work also highlights the stiffening of the EoS in the presence of light DM, thereby satisfying the maximum mass constraint from PSR J0740+6620, which was not satisfied by the NJL model alone for SQM with a proper choice of $G_V$. It is important to mention that the choice of the values of $m_{\chi}$ and the corresponding $\Lambda_{SqD}$ and $\Lambda_{VqD}$ is quite conservative (our SGP parameterization) because most of the parameter spaces for both the scalar and vector types of interaction are excluded by the various direct and indirect detection experiments for DM and the relic abundance bound. We also find that the distribution of DM in the star, controlled by the parameter $\alpha$, plays an important role in the DM population. This in turn affects the mass and radius of the DMSQSs. For fixed values of $G_V/G_S$ and $m_{\chi}$, a higher value of $\alpha$ (larger DM fraction) yields a more massive DMSQS with a larger radius. With proper choice of $\alpha$, we obtain DMSQS configurations that satisfy all the present day astrophysical constraints on the structural properties of compact stars.

We also study the non-radial $f$-mode oscillation of the DMSQSs with both Cowling approximation and GR treatment. Irrespective of the values of $m_{\chi}$, $\alpha$, and $G_V/G_S$, the universality related to $f_f$ and $\bar{\rho}$, $C$, and $\Lambda$, holds tight. These universal relations also do not depend on the composition of the compact star. We also find that our fitted relations for $f_f-\bar{\rho}$, $f_f-C$, and $f_f-{\rm{ln}}(\Lambda)$ are comparable with that obtained by the various other works considering different compositions of the compact star matter. Additionally, the relations $f_{f_{1.4}}-R_{1.4}$ and $f_{f_{1.4}}-\Lambda_{1.4}$ show negative correlations and universality regardless of the values of the free parameters of the DNJL model. The measurement of the oscillation frequency depends on the sensitivity of the upcoming GW detectors like the aLIGO, A+, CE1, and ET. We examined the detectability of $f_f$ by calculating the amplitude of the GW strain ($h_+$) assuming that pulsars of mass between (1 - 2.25)$M_{\odot}$ exist as DMSQS in the Messier 3 (M3 or NGC 5272) globular cluster at a distance of 10 kpc. Under such conditions, $f_f$ can be measured by all the four GW detectors. Our analysis shows that detection of $f$-mode oscillation may be possible for pulsars at distance (1$-$10) kpc, emitting GW radiation energy in the range $\sim10^{42}-10^{44}$ ergs.

Our DNJL model has three free parameters $G_V/G_S$, $\alpha$, and $\rho_{sc}$, which are optimized by performing Bayesian analysis using several astrophysical constraints on the structural properties of the compact stars. The results set the most credible ranges of the three parameters for $m_{\chi}=100~(150)~\rm{MeV}$ as $G_V/G_S = 0.398^{+0.332}_{-0.119}~ (0.414^{+0.329}_{-0.133})$, $\alpha = 0.194^{+0.029}_{-0.080}~ (0.198^{+0.038}_{-0.084})$, and $\rho_{sc} = 1.007^{+0.382}_{-0.268}~ (1.005^{+0.388}_{-0.348})~\rm{fm^{-3}}$. Interestingly, we note that the optimized parameter set remains nearly independent of the DM mass and the momentum cut-offs of the dark sector.


\section*{Acknowledgements}
Work of D.S. is supported by the NRF research Grant (No. 2023R1A2C1003177). Work of A.G. and J.C.P. is supported by the National Research Foundation of Korea (MSIT) (RS-2024-00356960).

\appendix
\section{Formalism of calculation of $f$-mode frequency}
\label{Sec:Appendix}

\subsection{Cowling approximation}
\label{sec:Cowling}

In order to calculate the oscillation frequencies of the non-radial oscillation of compact stars, a widely used approximation is known as the Cowling approximation which ignores the metric perturbations. Once the TOV Eqs. (\ref{eq:tov} - \ref{eq:tov3}) are solved, we obtain the oscillation mode frequencies by solving the two following coupled differential equations:
\begin{eqnarray}
\frac{dW(r)}{dr}&=&\frac{d\varepsilon(r)}{dP(r)}\Bigg[\omega^2r^2e^{\lambda(r)-2\Phi(r)}V(r) + \frac{d\Phi(r)}{dr}W(r)\Bigg]- l(l+1)e^{\lambda(r)}V(r),
\label{W eqn}\\
\frac{dV(r)}{dr}&=&2\frac{d\Phi(r)}{dr}V(r) - e^{\lambda(r)}\frac{W(r)}{r^2}.
\label{V eqn}
\end{eqnarray}
For solving the Eqs. (\ref{W eqn}) and (\ref{V eqn}), the two boundary conditions are imposed at the center ($r=0$) and the surface ($r=R$) of the star. Near the center ($r=0$) of the star, the functions $W(r)$ and $V(r)$ behave as follows.
\begin{eqnarray}
W(r)=Ar^{l+1}~{\rm{and}}~~V(r)=-Ar^l/l;
\label{bc_center}
\end{eqnarray}
where, $A$ is an arbitrary constant. Toward the surface ($r=R$) of the star, another boundary condition for $W(r)$ and $V(r)$ is obtained as
\begin{eqnarray}
\omega^2R^2e^{\lambda(R)-2\Phi(R)}V(R) + \frac{d\Phi(r)}{dr}\Bigg\rvert_{r=R}W(R) = 0.
\label{bc_surface}
\end{eqnarray}
By assuming an initial value of $\omega^2$, we integrate the coupled differential equations, Eqs.\ref{W eqn} and \ref{V eqn} from the center to the surface of the star. After each integration, we use the Ridders' method to improve the value of $\omega^2$ until Eq. (\ref{bc_surface}) is satisfied. 

\subsection{General relativistic approach}
\label{sec:GR}
The line element for the perturbed Regge-Wheeler metric is given as
\begin{eqnarray}
\begin{aligned}
ds^{2}= & - e^{2\Phi}(1+r^{l}H_{0}Y_{m}^{l} e^{i \omega t})dt^{2} -2 i \omega r^{l+1}H_{1}Y_{m}^{l} e^{ i \omega t} dt dr + e^{2\lambda}(1-r^{l}H_{0}Y_{m}^{l} e^{ i \omega t}) dr^{2} \\
 & + r^{2}(1-r^{l}KY_{m}^{l} e^{ i \omega t})( d \theta^{2}+\sin^{2}\theta d\phi^{2}).
\end{aligned}
\end{eqnarray}
Here, $H_0, H_1,$ and $K$ are the metric perturbation functions; $\omega$ is the complex oscillation frequency. On the other hand, the perturbation of the fluid inside the star is quantified by the Lagrangian displacement vector $\boldsymbol{\xi} \left( \xi^{r}, \xi^{\theta}, \xi^{\phi} \right)$, which is defined in terms of the amplitudes of the perturbation as $W(r)$ and $V(r)$ follows,
\begin{eqnarray}
\begin{aligned}
 & \xi^{r}=r^{l-1} e^{-\lambda}W(r)Y_{m}^{l} e^{i \omega t}, \\
 & \xi^{\theta}=-r^{l-2}V(r)\partial_{\theta}Y_{m}^{l} e^{i \omega t}, \\
 & \xi^{\phi}=-r^{l}(r\sin\theta)^{-2}V(r)\partial_{\phi}Y_{m}^{l} e^{i \omega t}.
\end{aligned}
\end{eqnarray}
An additional fluid perturbation function $X$ has been introduced in \cite{Lindblom:1983ps, Detweiler:1985zz} for computational convenience, which is related to the Lagrangian pressure variations by the relation,
\begin{eqnarray}
\Delta P = - r^l e^{-\Phi} X Y^l_{m} e^{i \omega t}. 
\end{eqnarray}
The perturbations of a spherical star can have four degrees of freedom. Two of the six perturbation functions can be expressed in terms of the other four by solving the perturbed Einstein equation, $\delta G^{\mu \nu} =8 \pi \delta T^{\mu \nu} $. Following the numerical method developed in \cite{Lindblom:1983ps, Detweiler:1985zz, Lu:2011zzd}, we choose the four independent variables as $K, H_1, W,$ and $ X $. The other two functions $H_0$ and $V$ are converted to functionals of $K, H_1, W,$ and $ X $ using the algebric relations as follows.
\begin{eqnarray}
\begin{aligned}
H_{0} & =\left\{8\pi r^{3} e^{-\Phi}X-\left[\frac{1}{2}l(l+1) \left(m+4\pi r^{3} P \right)-\omega^{2}r^{3} e^{-2(\lambda+\Phi)}\right]H_{1}\right. + \bigg[\frac{1}{2}(l+2)(l-1)r-\omega^2 r^3 e^{-2\Phi} \\
& - \frac{e^{2\lambda}}{r} \left(m+4\pi r^{3} P \right) \left(3m-r+4\pi r^{3} P \right)\bigg] K\bigg\} \times\left\{3m+\frac{1}{2}(l+2)(l-1)r+4\pi r^3P\right\}^{-1}, \\
V & =\left\{\frac{X}{\varepsilon + P}+\frac{1}{r}\frac{dP}{dr} e^{\Phi-\lambda} \frac{W}{\varepsilon + P} - \frac{1}{2} e^{\Phi}H_{0}\right\}  \times \frac{e^{\Phi}}{\omega^{2}}. 
\end{aligned}
\end{eqnarray}
The homogeneous linear differential equations for the independent perturbation functions
\begin{eqnarray}
\begin{aligned}
 & r \frac{dK}{dr} = H_{0}+\frac{1}{2}l(l+1)H_{1}-\left[(l+1)- r \frac{d\Phi}{dr}\right]K - 8\pi(\varepsilon + P) e^{\lambda} W, \\
& r \frac{dH_1}{dr} = -\left[l+1+ \frac{2m}{r} e^{2\lambda}+4\pi r^{2} e^{2\lambda}(P-\varepsilon)\right]H_{1}+ e^{2\lambda}\left[H_{0}+K\right], \\
& r \frac{dW}{dr} = -(l+1) W + r^2 e^{\lambda}\Big[(\gamma P)^{-1} e^{-\Phi}X-\frac{l(l+1)}{r^{2}} V+\frac{1}{2}H_{0}+K\Big], \\
& r \frac{dX}{dr} = -l X+ (\varepsilon + P) e^{\Phi}\biggl\{\frac{1}{2}(1- r \frac{d\Phi}{dr})H_{0}+\frac{1}{2}\biggl[r^2 \omega^{2} e^{-2\Phi}+\frac{1}{2}l(l+1)\biggr]H_{1}+\frac{1}{2}(3 r \frac{d\Phi}{dr}- 1)K \\
& - \frac{l(l+1)}{r} \frac{d\Phi}{dr} V-\Big[4\pi(\varepsilon + P) e^{\lambda}+\omega^{2}e^{\lambda-2\Phi}-r^{2} \frac{d}{dr} \left(\frac{e^{-\lambda}}{r^{2}} \frac{d\Phi}{dr}\right)\Big]W\biggr\}.
\end{aligned}
\label{Eq:Fourth_order_diff_eqn}
\end{eqnarray}
Here, $\gamma$ is the adiabatic index and is defined as $\gamma = \frac{(\varepsilon + P)}{P} \frac{dP}{d \varepsilon}$. For each $l$ and $\omega $ there exist four linearly independent solutions for these four coupled linear differential equations without any boundary condition imposed. We define the system of perturbation functions as $Y(r) = \left\lbrace H_1, K, W, X \right\rbrace $. The system is singular at $ r = 0 $. For solutions near the center of the star, the perturbation functions are expanded in the power of $r$.
\begin{eqnarray}
Y(r) = Y(0) + \frac{1}{2} Y_{,rr}(0) r^2 + \mathcal{O}(r^4).
\end{eqnarray}
The relevant relations for the higher order terms of this approximation are taken from \cite{Lu:2011zzd}. The boundary conditions are imposed to ensure that the functions are finite everywhere inside the star. Also, at the surface of the star, the perturbed pressure should be zero for a consistent choice of $X$. The lowest order terms satisfy the relations at $ r = 0 $
\begin{eqnarray}
\begin{aligned}
 & H_{0}(0)= K(0),  \\
 & H_{1}(0)= \frac{1}{l(l+1)}\big[2lK(0)+16\pi(\varepsilon_{0}+P_{0})W(0)\big],  \\
 & X(0)= (\varepsilon_{0} + P_{0} ) e^{\Phi_0}\biggl\{\left[\frac{4\pi}{3}(\varepsilon_0+3P_0)-\omega^2 e^{-2\Phi_0}l^{-1}\right] \times W(0)+\frac{1}{2}K(0)\biggr\}. 
\end{aligned}
\label{Eq:first_order_relations}
\end{eqnarray}
Here, $\varepsilon_{0}$, $P_{0}$, and $\Phi_0$ represent the values of the corresponding variables at $r=0$. While integrating the system of differential equations, Eq.(\ref{Eq:Fourth_order_diff_eqn}), starting at $ r = 0 $, we impose the boundary conditions \cite{Lindblom:1983ps, Detweiler:1985zz, Lu:2011zzd, Guha:2025ssq} as follows.
\begin{eqnarray}
W(0) = 1, ~ K(0) = \pm (\varepsilon_0 + P_0).
\end{eqnarray}
$H_1(0)$ and $X(0)$ are calculated using Eq.(\ref{Eq:first_order_relations}) for these two choices. Clearly, there are two independent solutions of Eq.(\ref{Eq:Fourth_order_diff_eqn}) while starting the integration from $r=0$. On the other hand, there is only one boundary condition, $X(R) = 0$, while starting the integration from $ r = R $. In that case, there exist three independent solutions and the boundary values of $H_1, K$ and $W$ are chosen arbitrarily. To avoid singularity we integrate Eq.(\ref{Eq:Fourth_order_diff_eqn}) starting from $ r = R_G $, where, $R_G$ is very close to $R$ \citep{Lu:2011zzd}. We choose one point inside the star, e.g., $ r_c = R/2 $, and perform integration of Eq.(\ref{Eq:Fourth_order_diff_eqn}) from $r_0$ to $r_c$ (forward) (where $r_0$ is very close to the center) and from $R_G$ to $r_c$ (backward). To satisfy the boundary conditions both at the center and at the surface of the star, the linear combination of the two independent forward solutions and the linear combination of the three independent backward solutions are matched. The corresponding coefficients of weights are solved for each of the independent solutions and the final solutions of Eq.(\ref{Eq:Fourth_order_diff_eqn}) are obtained which is valid everywhere inside the star. Outside the star, the fluid perturbation functions vanish, i.e. $W = X = 0$. Eq.(\ref{Eq:Fourth_order_diff_eqn}) reduces to a second order system which is named as the Zerilli equation \cite{Lindblom:1983ps, Detweiler:1985zz, Lu:2011zzd, Guha:2025ssq}
\begin{eqnarray}
\frac{d^2 Z}{dr^{*2}}+
\begin{bmatrix}
\omega^2-V(r^*)
\end{bmatrix}Z=0,
\label{Eq:Zerilli}
\end{eqnarray}
where, $V(r^*)$ is the effective potential and is given by
\begin{eqnarray}
\begin{aligned}
V(r^*) & =\frac{2(1-2m/r)}{r^3(nr+3m)^2}[n^2(n+1)r^3 +3n^2mr^2+9nm^2r+9m^3].
\end{aligned}
\end{eqnarray}
$r^*$ is the tortoise coordinate and is given in terms of $r$ as
\begin{eqnarray}
r^* = r + 2 m \log \left( \frac{r}{2 m} - 1 \right),
\end{eqnarray}
and $n = (l-1)(l+2)/2$. In terms of $H_0(r)$ and $K(r)$ outside the star, the newly defined Zerilli function and its first derivative are defined as 
\begin{eqnarray}
\begin{aligned}
Z(r^{*}) &= \frac{k(r)K(r)-a(r)H_{0}(r)-b(r)K(r)}{k(r)g(r)-h(r)}, \\
\frac{dZ(r^*)}{dr^*} &= \frac{h(r)K(r)-a(r)g(r)H_0(r)-b(r)g(r)K(r)}{h(r)-k(r)g(r)},
\end{aligned}
\end{eqnarray}
where, the radial functions are defined as
\begin{eqnarray}
\begin{aligned}
& a(r)=-(nr+3m)/\left[\omega^{2}r^{2}-(n+1)m/r\right], \\
& b(r)=\frac{\left[nr(r-2m)-\omega^{2}r^{4}+m(r-3m)\right]}{(r-2m)\left[\omega^{2}r^{2}-(n+1)m/r\right]}, \\
& g(r)=\frac{\left[n(n+1)r^{2}+3nmr+6m^{2}\right]}{r^{2}(nr+3m)}, \\
& h(r)=\frac{\left[-nr^{2}+3nmr+3m^{2}\right]}{(r-2m)(nr+3m)}, \\
& k(r)=-r^{2}/(r-2m).
\end{aligned}
\end{eqnarray}
Now clearly the Zerilli equation has two independent solutions which represent incoming and outgoing waves, $Z_{+}(r*)$ and $Z_{-}(r*)$, respectively. The linear combination of these two represents the general solution as follows.
\begin{eqnarray}
Z(r^*)=A(\omega)Z_-(r^*)+B(\omega)Z_+(r^*).
\label{Eq:mode_expansion}
\end{eqnarray}
At very large $r$, the expansion of $Z_-$ and $Z_+$ can be represented as
\begin{eqnarray}
Z_{-}(r^{*})= e^{-i \omega r^{*}}\sum_{j=0}^{\infty}\beta_{j}r^{-j},~ ~ Z_{+}(r^{*})= e^{i \omega r^{*}}\sum_{j=0}^{\infty}\bar{\beta}_{j}r^{-j},
\label{Eq:expansion}
\end{eqnarray}
where, $\bar{\beta}_{j}$ is the complex conjugate of ${\beta}_{j}$. We numerically solve the Zerilli equation, Eq.(\ref{Eq:Zerilli}) for $r \leq 50 \omega^{-1}$. Now, at that large radius, keeping the terms up to second order, we get
\begin{eqnarray}
\begin{aligned}
& Z_{-}={e}^{-{i}\omega r^{*}}\left[\beta_{0}+\frac{\beta_{1}}{r}+\frac{\beta_{2}}{r^{2}}+{\mathcal{O}}(r^{3})\right], \\
& \frac{{d}Z_{-}}{\mathrm{d}r^{*}}=-{i}\omega{e}^{-{i}\omega r^{*}}\left[\beta_{0}+\frac{\beta_{1}}{r}+\frac{\beta_{2}-i\beta_{1}(1-2m/r)/\omega}{r^{2}}\right].
\end{aligned}
\label{Eq:substitution}
\end{eqnarray}
Substituting Eq.(\ref{Eq:substitution}) into Eq.(\ref{Eq:Zerilli}), we obtain \cite{Guha:2025ssq, Lu:2011zzd, Zhao:2022tcw}
\begin{eqnarray}
\begin{aligned}
 & \beta_{1}=\frac{-i(n+1)\beta_{0}}{\omega}, \\
 & \beta_{2}=\frac{[-n(n+1)+im\omega(3/2+3/n)]\beta_{0}}{2\omega^{2}}.
\end{aligned}
\end{eqnarray}
For the real values of $\omega$, the incoming and outgoing wave amplitudes are complex conjugates of each other, that is,
\begin{eqnarray}
A(\omega) = B^{*}(\omega).
\end{eqnarray}
At $r = 50 \omega^{-1}$, we obtain the numerical values of $Z$ and $\frac{dZ}{dr^*}$ by solving the Zerilli equation. Now using these numerical values, we can estimate $B(\omega)$ from Eq.(\ref{Eq:mode_expansion}) and its derivative. $B(\omega)$ is in general complex for each $\omega$. 

Next in the final step, we approximately interpolate $B(\omega)$ as an analytic function of $\omega$ along the real axis. Omitting the higher order terms we assume the power series as
\begin{eqnarray}
B(\omega) \approx \gamma_0 + \gamma_1 \omega + \gamma_2 \omega^2.
\end{eqnarray}
For the three guess values of the real part of $\omega$, we estimate $B(\omega)$ numerically. In the absence of the incoming wave, the oscillation frequency is determined as the solution of the three linear equations containing three unknown variables, $ \gamma_0, \gamma_1  $ and $ \gamma_2 $. The oscillation frequency is the root of   
\begin{eqnarray}
B(\omega) = \gamma_0 + \gamma_1 \omega + \gamma_2 \omega^2 = 0,
\end{eqnarray}
with a positive imaginary part. We repeat this process with different guess values (three guess values close to the previous solution) of the real $\omega$ until the solution converges to the eighth digit. The complex eigen frequency can be written as $\omega=2\pi f + i/\tau$, where $\tau$ is the damping time of the GW.


\bibliographystyle{apsrev4-1.bst}
\bibliography{ref}

\end{document}